\documentclass[twocolumn]{emulateapj}

\usepackage{amsmath}
\usepackage[title]{appendix}

\shorttitle{Young Cluster Age Analysis}
\shortauthors{Cummings et~al.}

\begin{document}

\title{Improved Main Sequence Turnoff Ages of Young Open Clusters: \\Multicolor UBV Techniques \& 
the Challenges of Rotation}

\author{Jeffrey D. Cummings\altaffilmark{1} \& Jason S. Kalirai\altaffilmark{2,1}}
\affil{}

\altaffiltext{1}{Center for Astrophysical Sciences, Johns Hopkins University,
3400 N. Charles Street, Baltimore, MD 21218, USA; jcummi19@jhu.edu}
\altaffiltext{2}{Space Telescope Science Institute, 3700 San Martin Drive, Baltimore, MD 21218, USA;
jkalirai@stsci.edu}

\begin{abstract}
Main sequence turnoff ages in young open clusters are complicated by turnoffs that are sparse, have high 
binarity fractions, can be affected by differential reddening, and typically include a number of peculiar 
stars.  Furthermore, stellar rotation can have a significant effect on a star's photometry and evolutionary 
timescale.  In this paper we analyze in 12 nearby open clusters, ranging from ages of 50 Myr to 350 Myr, how broadband 
UBV color-color relations can be used to identify turnoff stars that are Be stars, blue stragglers, certain 
types of binaries, or those affected by differential reddening.  This UBV color-color analysis also directly
measures a cluster's E(B--V) and estimates its [Fe/H].  The turnoff stars unaffected by these peculiarities
create a narrower and more clearly defined cluster turnoff.  Using four common isochronal models, two of which 
consider rotation, we fit cluster parameters using these selected turnoff stars and the main sequence.  Comparisons 
of the photometrically fit cluster distances to those based on Gaia DR2 parallaxes find that they are 
consistent for all clusters.  For older ($>$100 Myr) clusters, like the Pleiades and the Hyades, comparisons 
to ages based on the lithium depletion boundary method finds that these cleaned turnoff ages agree to within $\sim$10\% for
all four isochronal models.  For younger clusters, however, only the Geneva models that consider 
rotation fit turnoff ages consistent with lithium-based ages, while the ages based on non-rotating isochrones 
quickly diverge to become 30\% to 80\% younger.  This illustrates the importance of rotation for 
deriving ages in the youngest ($<$100 Myr) clusters.

\end{abstract}

\section{Introduction}

Open clusters provide an ideal environment to study star formation, stellar physics, and nearly 
all aspects of stellar evolution.  This is driven by their large sample of stars being at effectively 
the same initial composition, age, distance, and reddening.  For characterizing processes of stellar 
evolution using clusters, one of the most important factors is cluster age.  This allows multiple 
clusters across different ages to be used to trace the evolution of, for example, stellar rotation rates 
(e.g., Kawaler 1988, Barnes 2003), lithium abundances (e.g., Cummings et~al.\ 2012, 2017), cluster 
metallicity or $\alpha$ abundances (e.g., Reddy et~al.\ 2016), 
and their populations of peculiar stars (e.g., Marco et~al.\ 2007).  The most straightforward method of 
analyzing cluster ages is simply through analysis of their highest mass stars that are beginning to evolve 
beyond the main sequence in the main sequence turnoff (hereafter MSTO).  Plotting the cluster stars in a 
color-magnitude diagram (hereafter CMD) and fitting the main sequence, the MSTO, and, when available, the 
giants with a theoretical isochrone provides a cluster age.

Multiple challenges exist, however, with deriving precise cluster ages from fitting the MSTO.  For 
example, the isochrones themselves are sensitive to the model assumptions, including how convection and 
overshoot are handled.  Next, fitting isochrones is dependent on adopted cluster reddening, cluster chemical 
compositions, and cluster distances.  Differential reddening and binarity can further compound the challenge 
of MSTO analysis.  One of the most complex challenges for MSTO analysis is stellar rotation.  The understanding of 
rotation and its models have improved (see the review in Maeder \& Meynet 2000 and subsequent work by the 
Geneva group and Choi et~al.\ 2016), but there remain limitations and inconsistencies between models.  Rotation 
affects nearly all aspects of stellar evolution, including the observed characteristics of stars.  MSTOs with 
stars spanning a large range of rotations and observation angles are significantly broadened (Georgy et~al.\ 2014).

These MSTO challenges are the most important in younger clusters ($\leq$200 Myr).  Their MSTOs
are defined by intermediate and higher mass stars, which are fewer in number, have high binarity fractions (Raboud 1996, Dunstall et~al.\ 2015), 
and have rotation rates spanning from slow to nearly critical rotation (Huang et~al.\ 2010).  Such rapidly 
rotating stars not only undergo the effects of rotation but can also expel material into circumstellar decretion 
disks.  These `Be' stars are typically B dwarfs and produce Balmer emission lines.  While they are very well 
studied objects, understanding their formation and evolution remains complex (see Rivinius et~al.\ 2013).  In 
general, however, they can either appear bright and blue or faint and red based on the angle of observation and the 
disk's contribution to or occultation of the star's light.  These Be stars further broaden MSTO photometry, 
but even after their removal the current rotational models alone still cannot explain the full widths of young 
cluster MSTOs (e.g., Milone et~al.\ 2015, Correnti et~al.\ 2017).

A method to determine young cluster ages that bypasses many of these challenges is to derive ages from the
rich sample of lower-mass cluster members.  For example, the lithium depletion boundary (hereafter LDB) method 
uses lithium (hereafter Li) abundances of low-mass M dwarfs (e.g., D'Antona \& Mazzitelli 1994, Jeffries \& 
Naylor 2001).  LDB ages are believed to be less sensitive to model assumptions (Soderblom et~al.\ 2014), but 
they still have moderate sensitivity to magnetic fields and surface activity (e.g., Somers \& Pinsonneault 2014).  
Additionally, the challenge of spectroscopically observing Li abundances in the lowest-mass members of a young 
cluster is difficult to overcome in all but the nearest and most well defined Galactic clusters.  However, these LDB ages 
provide a valuable independent check of MSTO ages and methods.  

In Cummings et~al.\ (2016; hereafter Paper I), we used Johnson UBV color-colors of young cluster to 
analyze their reddening and metallicity, as is traditionally done, but we also used color-color analysis to 
identify higher-mass MSTO stars unaffected by peculiarities or differential reddening.  The CMD analysis of 
these selected MSTO stars provided a uniformly analyzed set of cluster ages.  These six clusters also had 
white dwarf members with spectroscopically determined masses and cooling ages, and combined with the cluster 
ages this gave each white dwarf's progenitor evolutionary timescale, and hence its progenitor mass.  These 
improved MSTO cluster ages gave an initial-final mass relation with observed scatter $\sim$50\% smaller than previous 
relations.  

Here we will improve this UBV color-color technique and more directly analyze which MSTO challenges it can 
mitigate, if not address, and reanalyze these six original clusters plus six additional young clusters.  An advantage 
of this technique is that Johnson broadband UBV photometry of cluster MSTO stars is relatively easy to acquire, if 
not already available, compared to other techniques to analyze peculiarities that require either narrow-band 
photometry or spectroscopy of MSTO stars.  More broadly speaking, while it will not be analyzed here, similar 
techniques can likely also be applied to color-color relations involving near-ultraviolet (NUV) filters like 
Sloan u and HST's F336W.

The layout of the paper is as follows.  In Section 2 we will describe four common model isochrones, and for 
the two that include rotation, we will look at how their predicted photometric effects of rotation compare.  
In Section 3 we will discuss 12 young open clusters and their UBV color-color diagrams, and use them to measure 
their reddening, estimate their metallicity, and identify peculiar types of higher-mass MSTO stars, in 
particular, those that could complicate age analysis.  In Section 4 we will discuss the fitting methods of these
cleaned cluster MSTOs, where these peculiar stars have been removed.  We will also look at how 
the differences in stellar models affect the derived ages and distance moduli.  In Section 5 we will 
compare the distance moduli of 10 of these 12 clusters, plus of 4 additional young and intermediate aged clusters, 
to those from the Gaia data release 2 (DR2; Gaia Collaboration et~al.\ 2016, 2018a).  We will also 
compare these MSTO ages to the LDB ages for five of these clusters.  In Section 6 we will summarize our results 
and conclusions.

\section{Stellar Models \& Isochrones}

In Paper I we photometrically analyzed six young clusters using both Yale-Yonsei isochrones (Yi 
et~al.\ 2001; hereafter Y$^2$ isochrones) and PARSEC isochrones 
(Bressan et~al.\ 2012) version 1.2S\footnote{Available at http://stev.oapd.inaf.it/cgi-bin/cmd}.  
In this paper we continue to use the original Y$^2$ isochrones and the newly updated PARSEC isochrones 
(version 1.2S+COLIBRI PR16 from Marigo et~al.\ 2017).  These updated PARSEC isochrones include
better post-main sequence evolution and improve the main sequence colors and matching.  

To expand the analysis of cluster parameters and to consider the important effects of rotation, we will 
also compare to the MIST isochrones (Dotter 2016, Choi et~al.\ 2016; version 1.1), which are based 
on Modules for Experiments in Stellar Astrophysics (MESA; Paxton et~al.\ 2011, 2013, 2015).  MIST has 
isochrones for both non-rotating stars and for stars initially rotating at 0.4 of their critical rotation 
velocity (hereafter $v_{crit}$).  Choi et~al.\ (2016) applies these initial rotations only for stars more massive than 
1.8 M$_\odot$, followed by a gradual slow down to zero rotation from 1.8 to 1.2 M$_\odot$.  Rotation in these 
lower-mass stars was not considered by Choi et~al.\ (2016) because they did not model magnetic braking, which 
becomes crucial for angular momentum loss in these lower masses ($<$1.2 M$_\odot$) with surface convection zones (Kraft 1967).  
For the higher-mass stars, the adoption of 0.4 of $v_{crit}$ is appropriate based on cluster B-star observations 
(Huang et~al.\ 2010), where there is a broad observed distribution spanning the full range of rotation rates 
and with a peak near 0.4 to 0.5 of v$_{crit}$.  To further consider the effects of rotation in the MIST models, 
Choi et~al.\ (2017) also give isochrones for stars initially rotating at 0.5 and 0.6 of $v_{crit}$, but they
only include stellar parameters and not UBV photometry. 

Driven by the differences in current rotational model predictions (see Section 2.1), here we also consider the 
Geneva stellar models (Ekstr{\"o}m et~al.\ 2012, Georgy et~al.\ 2013, 2014; hereafter the SYCLIST models).  The 
SYCLIST models model initial rotation rates from non-rotating to nearly critical rotation for stars from 1.7 
to 15 M$_\odot$.  For the full stellar mass range they model non-rotating stars and those with initial rotations 
at 0.568 of their critical angular momentum, which is equivalent to 0.4 of $v_{crit}$ based on the relation 
defined in Ekstr{\"o}m et~al.\ (2008).  These SYCLIST models are important for considering the complete range of 
rotation's potential effects, but the available isochrones only have 3 metallicities with no available interpolation 
at Z=0.002, 0.006, and 0.014 (Z=0.014 is their adopted solar Z).  This limits our ability to apply isochrones 
of appropriate metallicity to the cluster analysis, and so we will use these for only
clusters that are at or very near solar metallicity.  

It is important to reiterate that for both MIST and SYCLIST models, these model rotation rates 
are initial rotation rates.  Both models consider appropriate angular momentum loss during stellar evolution and 
its consequences, and we refer the reader to Choi et~al.\ (2016) and Georgy et~al.\ (2013), respectively, for 
detailed discussions of how each rotational model accounts for this important factor.
 
Comparing the detailed physics of all of these models is beyond the scope of this paper, but it is important to 
note how these models scale metallicity.  Methods of deriving cluster [Fe/H], both spectroscopic and photometric, 
are all relative and insensitive to what solar composition (hereafter Z$_\odot$) is adopted, but stellar isochrones 
use absolute Z-scale abundances.  The Y$^2$ isochrones adopt Z$_\odot$=0.018 from Grevesse and Sauval (1998), 
while the PARSEC models adopt Z$_\odot$=0.0152, the MIST models adopt Z$_\odot$=0.0142, the SYCLIST models adopt 
Z$_\odot$=0.014.  Here we apply the same 
[Fe/H] to all isochrones matched to a given cluster, but we scale each isochrone's Z abundance according to their 
adopted Z$_\odot$.  Like with [Fe/H], the color-color determination of cluster reddening (E(B--V)) is also 
insensitive to adopted composition, and we similarly apply the same reddening to all isochrones matched to a 
given cluster.  

\subsection{Effects of Rotation on Stellar Isochrones}

Both the MIST and SYCLIST stellar evolutionary models account for rotation and angular momentum loss
from the pre-main sequence to the asymptotic giant branch, if not beyond for a subset of MIST isochrones.  
The effects of rotation remain challenging to model, but it has been known to play an important role 
in nearly all aspects of stellar evolution (see Maeder \& Maynet 2000).  The magnitude 
of these effects remains sensitive to many factors, including the adopted efficiency of various types of 
rotational mixing and loss of angular momentum, which still remain poorly understood.  We will consider 
throughout this paper the effects of rotation based on the MIST models, but we will also
reference the SYCLIST models, which adopt more efficient rotational mixing and have typically
stronger rotational effects.  

\begin{figure}[!ht]
\begin{center}
\includegraphics[clip, scale=0.44]{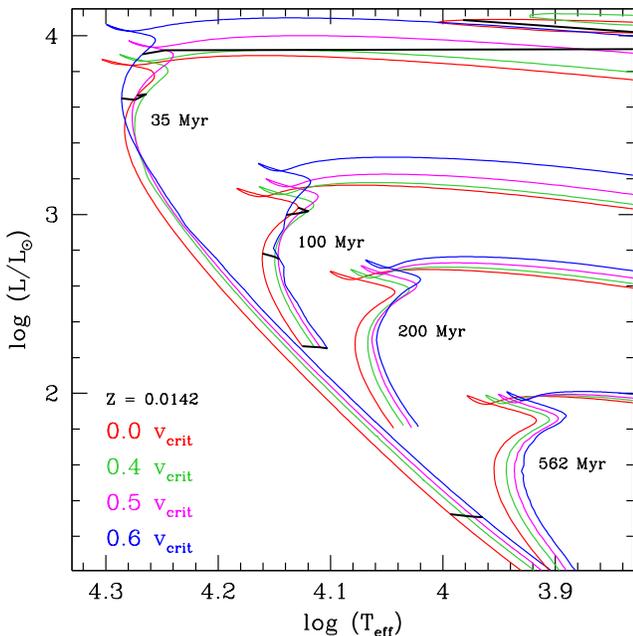}
\end{center}
\vspace{-0.4cm}
\caption{MIST Isochrones at various ages for non-rotating stars and stars initially rotating at 0.4, 
0.5, and 0.6 of $v_{crit}$  Solid black lines connect constant masses across the multiple
rotational velocities at a given age.  In the main sequence, radius inflation makes faster rotating stars
cooler, but in the MSTOs the slower evolution of the faster rotators makes them progressively bluer.
In all but the youngest ($<$100 Myr) MSTOs the MIST isochrones predict that the slowly (non) rotating
stars will define the blue edge of the MSTO.}
\end{figure}

\begin{figure}[!ht]
\begin{center}
\includegraphics[clip, scale=0.44]{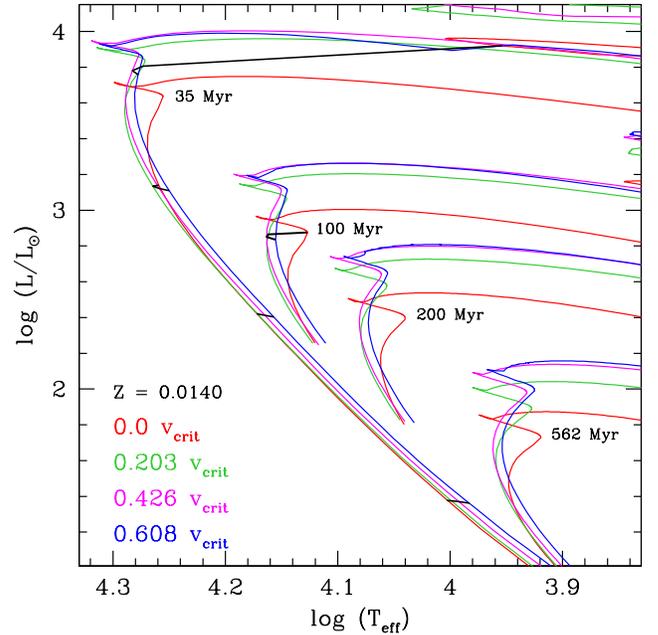}
\end{center}
\vspace{-0.4cm}
\caption{Similar to Figure 1, but comparing SYCLIST isochrones at constant ages over multiple initial
rotation rates.  The main sequences appear similarly affected by rotation in comparison to MIST models.  
The SYCLIST's models have more efficient rotational mixing, however, which leads to a stronger 
effect on the evolutionary rate.  For all cluster ages considered, the SYCLIST models predict that the moderate 
rotating stars (0.426 of $v_{crit}$) will define the blue edge of the MSTO.} 
\end{figure}

In Figure 1 we show MIST isochrones at ages of 35, 100, 200, and 562 Myr with no rotation and 
with initial rotation rates of 0.4, 0.5, and 0.6 of $v_{crit}$\footnote{For reference to the SYCLIST models, 
they define their rotations in terms of angular momentum and not equatorial velocity.  Based on the relation 
derived in Ekstr{\"o}m et~al.\ (2008), a rotation of 0.6 of $v_{crit}$ is approximately 0.8 of the critical angular 
momentum of a star.}.  Rotation produces two generally competing factors in a uniform-age MSTO.  First, rotation 
inflates the equatorial radius of a star, making it cooler but primarily conserving its total luminosity.  This 
is seen throughout the main sequence and the MSTO.  Second, rotation extends the lifetime of a star due to mixing 
additional hydrogen into its core.  This does not significantly affect the main sequence colors, but 
as stars approach the MSTO at a given mass the slowest rotators begin to evolve redward while the fastest 
rotators remain bluer and on the main sequence longer.  In Figure 1 this is seen most clearly at 35 Myr.  
When increasing from non-rotating to 0.4 of $v_{crit}$, the MSTO first becomes cooler because of radius
inflation, but at these higher masses for rotations of 0.5 to 0.6 of $v_{crit}$ the changes in evolutionary 
timescale begin to dominate and even with inflation they remain bluer.  In Figure 1 we help 
illustrate these effects by drawing black lines connecting several equal mass examples across the MIST 
isochrones at 35 Myr and 100 Myr.  For isochrones at 100 Myr and older the blue edge of the main 
sequence and MSTO model distribution are defined well by the non-rotating isochrones, but the MSTOs are 
more complex at younger ages.

For rotating stars, the magnitude of radius inflation is relatively insensitive to the details of the rotational 
models.  How much rotation affects the evolutionary timescale of a star, however, is very sensitive to the model's 
adopted magnitude of rotational mixing.  For comparison, Figure 2 shows a series of SYCLIST isochrones across a 
similar range of initial rotation rates, and for a given rotation these adopt more mixing.  Hence, the main sequence 
variations appear comparable to those of the MIST models, but the SYCLIST MSTOs become more sensitive to rotation 
(Georgy et~al.\ 2014).  At the MSTOs for all ages the isochrones with rotation all deviate significantly from the 
non-rotating isochrone, and in these models the 0.426 of $v_{crit}$ (0.6 of $\Omega_{crit}$) isochrone defines 
well the blue edge of each MSTO's model distribution.  To again help illustrate rotation's effects, we draw a series 
of solid black lines connecting several equal-mass examples across isochrones at a given age.  

Lastly, the luminosities and T$_{\rm eff}$ plotted in Figure 1 and 2 are total output luminosity and
mean T$_{\rm eff}$.  With increasing rotation, a star becomes increasingly oblate, and with 
gravitational darkening, there are increasing variations of T$_{\rm eff}$ across its surface with the poles
being the hottest and the equator being the coolest.  Therefore, 
a star's observed color and magnitude become increasingly sensitive to the observed inclination angle (see 
Townsend et~al.\ 2004, Georgy et~al.\ 2014).  Below rotations of 0.5 of $v_{crit}$ these effects are not
typically important, but they become moderate for stars at rotations from 0.5 to 0.6 of $v_{crit}$, 
especially for the extreme cases of 0$^{\rm o}$ and 90$^{\rm o}$ angles.  At rotation rates of 0.7 of $v_{crit}$ 
and faster, the effects of observation angle become significant.

\begin{figure}[!ht]
\begin{center}
\includegraphics[clip, scale=0.44]{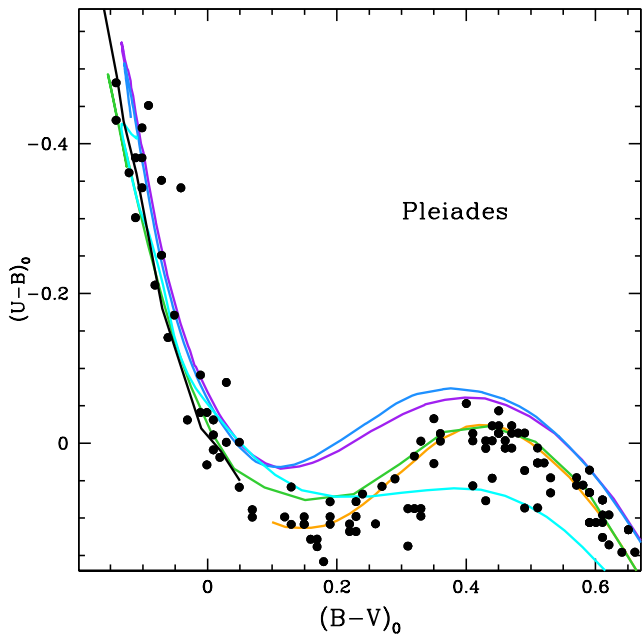}
\end{center}
\vspace{-0.4cm}
\caption{Pleiades color-color diagram with reddening corrected data plotted versus the
metallicity corrected (to [Fe/H]=+0.01) Hyades fiducial (orange), the Y$^2$ isochrone (green), the PARSEC
isochrone (purple), the MIST isochrone (blue), the SYCLIST isochrone (cyan), and the F70 empirical relation
(black).}
\end{figure}

\section{Color-Color Analysis}

In Paper I we considered preliminary color-color (U--B versus B--V) techniques to address the challenges 
of higher-mass MSTOs and to identify likely problematic stars.  Figure 3 shows the color-color diagram for 
the Pleiades MSTO and main sequence stars with photometry from Johnson \& Mitchell (1958).  Only
stars with Gaia DR2 parallaxes and proper motions consistent with Pleiades membership are displayed.
When plotting young cluster MSTO stars in color-color space ((B--V)$_0$ $<$ 0.1) they show a generally linear 
and consistent trend but also exhibit stars that deviate from this ``normal" trend.  Here we will analyze what 
these deviations may represent.

First, these hotter MSTO stars ((B--V)$_0$ $\leq$ 0.1) have a U--B versus B--V relation that is not 
sensitive to metallicity, unlike stars at redder colors.  This results from three factors, first, at hotter 
temperatures the bound-free and free-free opacities dominate here rather than the more metal sensitive H$^-$ 
opacity.  Second, the U band is full of metal lines in cooler stars, which causes strong line blanketing,
but this is less important at these hotter temperatures.  Third, while more metal-rich stars here still become 
marginally redder in B--V, they also become redder in U--B in a manner that shifts them along, rather than away from the U--B 
versus B--V relation (see the illustration of this in Figure 2 of Paper I).  Furthermore, as a cluster ages the 
position of the color-color trend does not evolve or change meaningfully for main sequence stars or even for 
subgiants (e.g., Fitzgerald 1970; hereafter F70).  When stars at the top of the MSTO evolve to the red in B--V 
they fold back to the red in U--B along the same color-color trend.  A cluster's age only affects how far its MSTO 
extends to the blue in U--B versus B--V.

Reddening is the only cluster parameter that affects the position of a young cluster's MSTO color-color relation.  
Therefore, the MSTO color-colors provide a reddening that is independent of a cluster's metallicity, age, and 
distance.  This reddening sensitivity will also cause MSTO stars that are affected by differential reddening 
across the field of an open cluster to deviate from the primary cluster color-color data.  In Figure 4 this 
color-color sensitivity to E(B--V) is illustrated.  However, differential reddening 
is not the only source causing stars to deviate in color-color.  We will discuss in Section 3.2 the effects on 
UBV color-color trends of binarity, stellar disks, chemically peculiar stars, rapid rotation, and classical blue 
stragglers, all of which are stars that can affect MSTO age fits.

\subsection{Empirical Versus Theoretical Relations}

Before we can analyze which stars have a normal color-color relation, we must define this relation.  Additionally, 
with the color-color relation's sensitivity to reddening, the method of applying the reddening to an empirical or 
theoretical reference is foundational to the cluster analysis.  
For E(B--V), like in Paper I, we again use its intrinsic dependence on (B--V)$_0$ from Schmidt-Kaler (1961) as 
described in Fernie (1963).  For E(U--B), in Paper I we simply adopted that E(U--B)=E(B--V)$\times$0.685.  Here we 
use the transformation from E(B--V) to E(U--B) from Crawford \& Madwewala (1976), using their version that is based 
on the interstellar absorption curves of Whitford (1958).  This more detailed transformation from E(B--V) to E(U--B) 
is comparable to our Paper I adoption but also incorporates a sensitivity to a star's intrinsic (B--V)$_0$.  This 
will have important effects on the analysis of clusters with moderate reddenings near E(B--V)=0.20.  We acknowledge 
that variations in these reddening relations may exist in different regions of the sky (e.g., Turner 1989), but all 
of the clusters analyzed here have low to moderate reddening where the consequences of these potential variations
are minimal.

When analyzing color-color relations, the well-studied Hyades fiducial is commonly adopted as an empirical Johnson 
UBV magnitude reference at a single age and metallicity (Figure 3; orange line).  It can be used to measure the 
reddening and estimate the cluster metallicity in lower-mass stars.  (See Paper I for the methods of measuring 
reddening and metallicity using the Hyades relation.)  The metallicity corrected Hyades fiducial, however, cannot 
be used to analyze the higher-mass MSTO stars in young clusters because it is an intermediate-aged cluster where these 
stars have fully evolved.  

The Y$^2$ color-color isochrone (Figure 3; green line) at intermediate- and low-masses ((B--V)$_0$ $>$ 0.3) 
closely matches the metallicity corrected Hyades fiducial.
For the Pleiades and other young clusters, matching the Y$^2$ model to the higher-mass MSTO stars finds an 
identical reddening to that derived with matching the Hyades fiducial to the lower-mass main sequence (see 
Paper I and its Figure 2).  Therefore, in Paper I we adopted the Y$^2$ isochrones to directly measured reddening
at the MSTO, insensitive to the adopted age and metallicity, and to identify the normal MSTO stars.  

In this paper we want to improve this technique to photometrically identify normal MSTO stars at 
these higher masses.  We first test whether we can use isochronal models consistent with our subsequent 
CMD isochronal age analysis (Section 4).  For the updated PARSEC isochrones (Figure 3; purple line), they 
follow the general observed trends, but the U magnitudes are too bright in the intermediate-mass stars and are 
marginally too bright in the MSTO stars.  This requires a lower reddening (E(B--V)=0.015
compared to E(B--V)=0.03) to match the Pleiades MSTO observations.  

Analyzing the MIST isochrone models (Figure 3; blue line) shows a similar color-color trend to that 
seen in the PARSEC models, with comparable inconsistencies with observations.  The MIST isochrones 
for stars initially rotating at 0.4 of $v_{crit}$ are not shown, but they are consistent at all 
colors other than a moderate difference for stars ranging from (B--V)$_0$ of 0.1 to 0.4.  The 
rotational sensitivity in this range results from the Balmer jump, which is in the U band 
and peaks in strength near a (B--V)$_0$ of 0.0.  From (B--V)$_0$ of 0.1 to 0.4 the Balmer jump has an 
important sensitivity to both temperature and surface gravity, and hence the color-color relation here 
is sensitive to rotation.  At higher T$_{\rm eff}$, however, where these young MSTO stars fall, the 
Balmer jump's sensitivity to surface gravity, and hence rotation, greatly diminishes.  More direct 
observational evidence that surface gravity plays no major role is that the 
F70 empirical color-color relations for subgiants and giants at this higher T$_{\rm eff}$ range 
are consistent with that of main sequence. 

Lastly, the SYCLIST model isochrones provide a complete range of initial rotations 
from non-rotating to 0.95 of $\Omega_{crit}$ (or 0.812 of $v_{crit}$).  Across this entire range of 
rotations, the MSTO model colors-colors are again not affected by rotation.  The general trends and 
fit reddenings are consistent with that of the Y$^2$ isochrones and the Pleiades observations, but for 
intermediate-mass stars the U magnitudes are far too bright and rapidly become too faint in the 
lower-mass stars.

The B and V magnitudes across all of these models are generally consistent, so these color-color 
differences illustrate the remaining challenges of synthesizing U magnitudes from stellar models. 
This also suggests that while neither these higher-mass 
MSTO models nor observations yet show evidence for rotation strongly affecting color-color relations, 
a more complete understanding of U magnitudes are necessary to explicitly state this.  For the rest of
this paper, however, we will adopt that the color-color relations of these MSTO are not affected
by rotation. 

To address these synthetic U challenges, F70 provides a solution because they give an empirical 
color-color relation derived from observations of nearby field stars.  Taking advantage of the 
color-color trends of these higher-mass stars being insensitive to distance, metallicity, age, and 
rotation, the variations of these parameters in field stars do not affect the empirical relation.  
Additionally, because the stars F70 used are relatively nearby field dwarfs, the effects of reddening 
are limited, but F70 still adopted reddening correction methods to increase their dereddened sample 
and improve statistics.  

In Figure 3 the F70 empirical relation matches well with the observed Pleiades and very well with 
the Y$^2$ isochrone at --0.07 $<$ (B--V)$_0$ $<$ +0.07, but the F70 relation at the bluest colors has 
a more rapid decrease in U--B with decreasing B--V.  This curvature is comparable to the curvature 
predicted in both the MIST and PARSEC isochrones.  These bluest stars are the most age sensitive and 
provide the best MSTO ages.  Therefore, the adoption of the empirical F70 relation and of more detailed 
reddening relations strongly improves the methods from Paper I.

\subsection{Color-Color Effects of Be Stars, Peculiar Abundances, Binarity \& Errors}

Differences in rotation, age, and composition are all commonly used to help explain the observed widths in young 
cluster MSTOs.  However, the color-color relations for such stars are comparably insensitive to all of these parameters, and 
their MSTOs still remain broadly distributed in this space.  Therefore, this provides a valuable tool to consider 
what other effects or peculiarities may be playing a role in broadening these complex young cluster MSTOs in both
color-color and CMD space.  

The first two peculiarities that we will consider are common in higher-mass MSTO stars: rapidly rotating Be stars 
and blue stragglers, which in young clusters are commonly but not always the same stars (e.g., Marco et~al.\ 2007).  
Be stars are well studied but remain challenging to fully explain.  The light contributed from their disks, which 
result from their rapid rotation, cause them to vary photometrically both optically and in the NUV in ways that 
make them deviate from the standard color-color trends (see discussions in Rivinius et~al.\ 2013).  The UBV 
color-color models of Sigut \& Patel (2013) can reproduce the generally observed variability and its dependence on 
inclination angle.  Be stars can go through phases of having increased NUV flux and phases of having decreased NUV 
flux relative to a normal B star (Zorec et~al.\ 1983).  The magnitude of NUV deviation from normal appears to correlate 
with the observed emission line strength (Harmanec 1983).  In general the NUV flux decreases occur in Be shell 
stars (those seen edge on) with the disk occulting light from the star and NUV flux increases occur in Be stars seen 
closer to pole on.  The variability may be explained by variations in the density of their disk, which dissipates and 
refills over time, and that a Be star can go through periods of having a weak disk and relatively normal colors and 
magnitudes.  

\begin{figure}[!ht]
\begin{center}
\includegraphics[clip, scale=0.44]{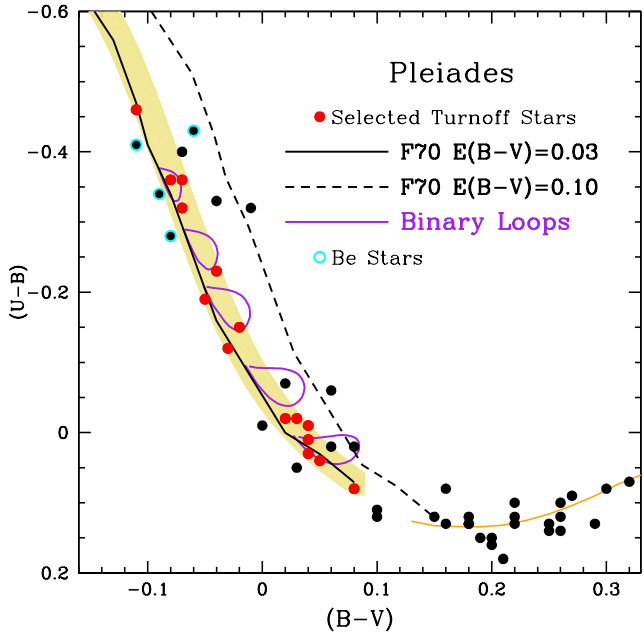}
\end{center}
\vspace{-0.4cm}
\caption{Pleiades color-color diagram, where in tan we show the F70 relation (solid black) based color-color range for 
our selected MSTO stars, which are shown as red data.  Be stars in the Pleiades are outlined
in cyan.  In purple we show how integrated binary color-colors vary and create loops for 5 example 
primaries with companions ranging from equal mass to the lowest mass stars.  Lastly, to illustrate the MSTO's
color-color sensitivity to E(B--V), we show our measured reddening at E(B--V)=0.03 using the F70 relation in solid black and 
E(B--V)=0.10 in dashed black.  Note that in this figure only we directly plot the observed colors and apply reddening
to the F70 relation and binary loops, which allows us to properly display more than one reddening relative to the observed 
data.}
\end{figure}

\begin{figure*}[!ht]
\begin{center}
\includegraphics[clip, scale=0.91]{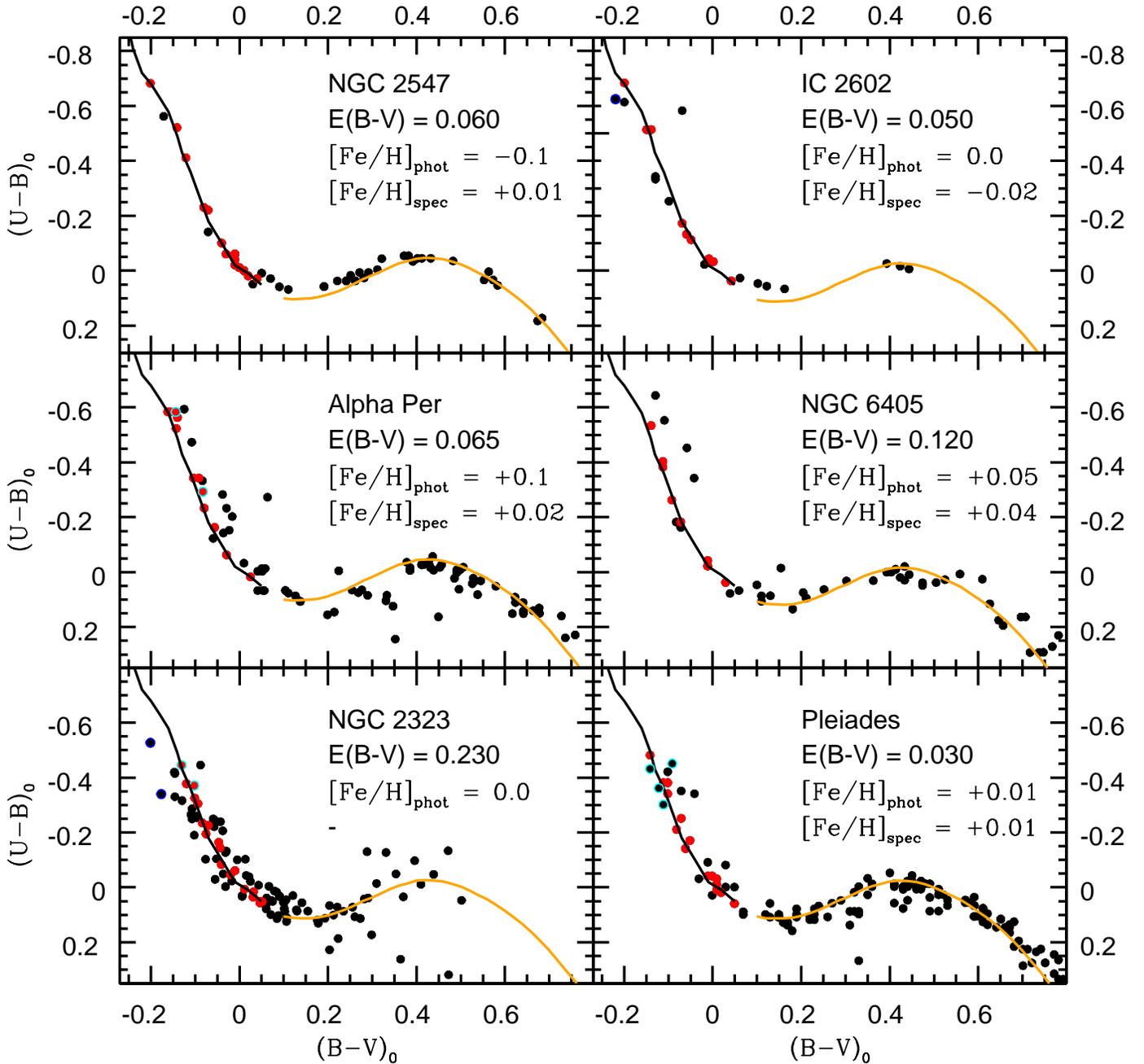}
\end{center}
\vspace{-0.4cm}
\caption{Color-color analysis of the six youngest open clusters in our sample.  The cluster displayed in each
panel is labeled.  The F70 relation fits are shown in black with the color-color selected MSTO stars in red.  The
Hyades fiducials, which have been metallicity corrected to match observations, are shown in orange.   Additionally, 
Be stars are circled in cyan and blue stragglers that are not Be stars are circled in blue.  See
Table 1 and Table 3 for the photometric sources and measured cluster parameters, respectively.}
\end{figure*}

\begin{figure*}[!ht]
\begin{center}
\includegraphics[clip, scale=0.91]{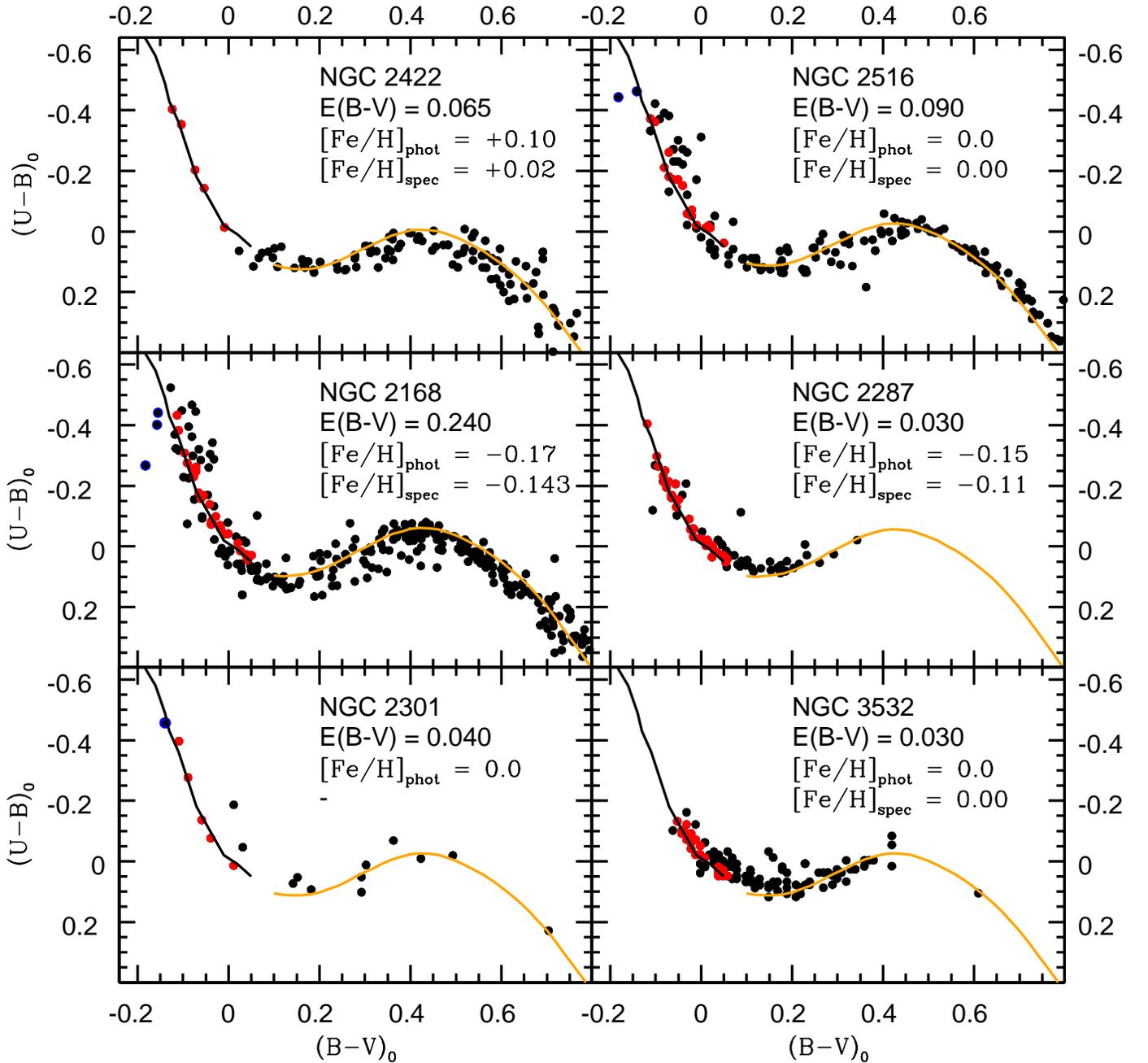}
\end{center}
\vspace{-0.4cm}
\caption{The same color-color analysis as in Figure 5 for the six older open clusters in our sample.}
\end{figure*}

\begin{figure}[!ht]
\begin{center}
\includegraphics[clip, scale=0.44]{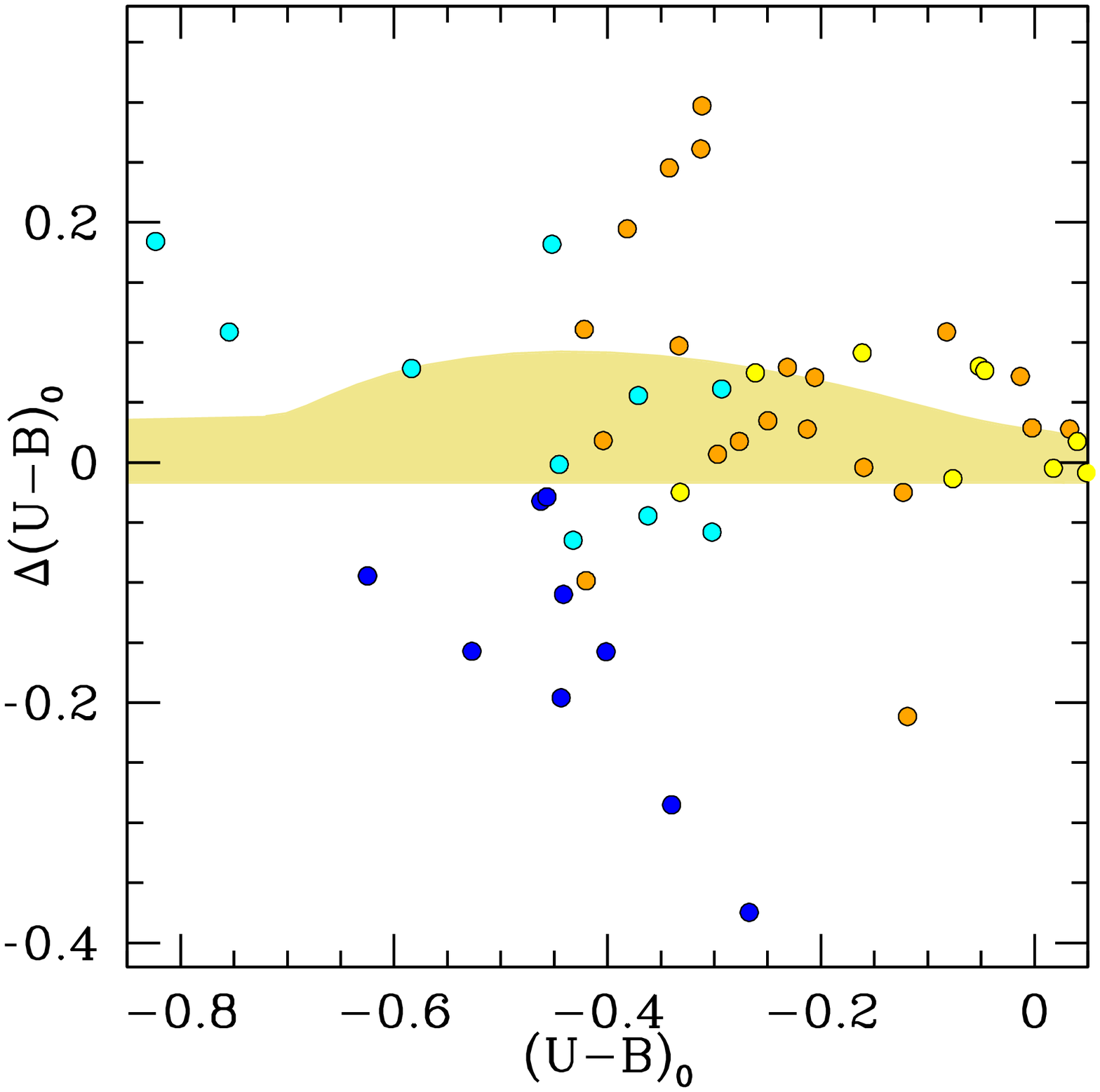}
\end{center}
\vspace{-0.4cm}
\caption{(U--B)$_0$ residuals in color-color space relative to the F70 relation.  The MSTO selected range
is again shown in tan, which in this residual space has a curvature due to how we follow the F70 relation 
in color-color space.  The peculiar stars with Gaia-based parameters consistent with membership
in the 12 clusters analyzed here are shown, with Be stars in cyan, blue stragglers that are not Be
stars in blue, peculiar abundance Am stars are shown in yellow, and Ap stars are shown in orange.}
\end{figure}

This large variability of Be stars is further illustrated by Ghosh et~al.\ (1999), with their spectroscopic 
re-observations of B stars originally defined as Bn stars (have strong rotational broadening but no 
emission).  They found during a short period of re-observations that $\sim$10\% of the Bn sample now do 
have emission and appear as normal Be stars.  Conversely, several well studied Be stars have been observed 
to undergo multi-year periods without emission (e.g., Bjorkman et~al.\ 2002, Miroshnichenko et~al.\ 2012).  
This illustrates that a star can be identified as a Be in earlier spectroscopic observations, but it may go 
through phases of otherwise normal colors and luminosities, and the opposite is also true.  This variability 
cautions against the comparison of spectroscopic identification of Be stars and their emission strengths from 
one epoch to photometric observations from another epoch.  After acknowledging this caveat, however, the
sample of spectroscopically identified Be stars in the 12 clusters in this paper can test if Be stars can 
generally be identified through their broadband color-color deviations alone.

For blue stragglers that are not identified as Be stars, the clusters studied here have a moderate sample (9)
of such stars.  While there is no clear understanding of how or why their color-color relations may differ
from ``normal" MSTO stars, if at all, our current analysis can test this.  If they uniformly appear 
abnormal, this will be a valuable insight.

B- and A-type dwarfs also commonly have peculiar surface abundances.  These can be driven by diffusive processes 
that affect slower rotators (e.g., Michaud 1970), where this slow rotation is typically caused by magnetic fields 
in Ap stars (St{\c e}pie{\'n} 2000) or binary interaction in Am stars.  Am and Ap stars have colors that typically 
make them appear more metal-rich.  Hence, like general metallicity variations, the color-color trends in the MSTO 
stars are not believed to be sensitive to these surface abundance peculiarities.  In early A dwarfs where the 
color-color trend plateaus and metallicity sensitivity becomes important, however, these peculiar abundances are common.  
This can partly explain why the color-color distribution significantly broadens in A dwarfs in most clusters.  These 
peculiar abundance stars systematically fall below the trend here with increasing line blanketing causing redder (U--B)$_0$ at 
comparable (B--V)$_0$ (St{\c e}pie{\'n} \& Muthsam 1980).  Fast rotating B and A dwarfs can also show peculiar 
abundances where rotational mixing enhances, for example, N and He at the surface (e.g., Maeder \& Meynet 2000).  
The SYCLIST models do not predict that these abundance peculiarities will affect the color-color trends themselves, 
but this likely needs further study.

The effects of binarity are more complex in color-color space than in CMD space.  In Figure 4, for primaries 
at five different masses/colors, we illustrate the effects of adding a companion ranging from an equal-mass 
companion to the lowest-mass companion.  We base this on the main sequence photometric relations of the Y$^2$ 
isochrones.  These binary color-color trends create a complete loop because equal-mass companions will have 
the same colors while the lowest-mass companions will have such low luminosity that the integrated colors will 
be dominated by the primary.  At the highest-mass primary considered, the effects are minor and mostly shift 
the integrated color-color along the normal relation and then back as the companion's luminosity decreases.  
With decreasing primary mass the companion's effects become more pronounced relative to the normal relation.  
This is because the color-color relation changes slope near a (B--V)$_0$ of 0.0.  If the primary is near this color, 
the secondary's luminosity will contribute more meaningfully to the binary's integrated colors as the secondary 
follows this slope change.

We define stars as consistent with the F70 relation if they are within the color-color band illustrated in 
Figure 4.  The range of this band is based on the distribution of MSTO stars for all clusters analyzed, in 
addition to the distribution of Be stars and blue stragglers.  The boundaries are created by shifting a 
smoothed F70 relation redward 0.017 magnitudes in (U--B)$_0$ and 
blueward 0.020 magnitudes in (B--V)$_0$.  This band helps to define a sharp and consistent blue edge to the 
selected MSTO, and shifting the relation along two axes helps to account for the curvature in the relation.  
We define the faint end of the color-color selected MSTO stars at (U--B)$_0$ = 0.06 and (B--V)$_0$ = 0.06 because 
here metallicity effects become important and the complexity of the A dwarfs dominate.  As seen in Figure 4 
these cuts will not be sensitive to all binaries but will remove those with the largest affected colors.  
In Figure 4 we also show the four known Pleiades Be stars as data points outlined in cyan, all of which fall 
outside of the selected range.  In Section 3.3 we will further analyze how this color-color range compares 
to previously identified Be stars, blue stragglers, and peculiar abundance stars in all 12 clusters. 

Lastly, these are strict color-color cuts on these MSTO stars, and we note that the effects of random photometric 
errors may bring some peculiar stars to within the range of acceptance or cause otherwise normal stars to be cut.  
This is a challenge for photometrically defining normal versus peculiar MSTO stars, but this identification is 
not our final goal, the goal is measuring cluster parameters.  We are using the same set of photometry for both the 
color-color and CMD analysis.  If a normal MSTO star has large enough photometric error to cause it to fall outside
the normal range and be removed during the color-color analysis, this error would also likely affect its CMD 
placement and potentially the cluster age fit.  Low photometric errors are ideal and the photoelectric and CCD 
photometry used for all of these clusters (see Table 1) provides this in these bright MSTO stars, but the effects 
of random photometric errors on the final MSTO fits are mitigated with these methods.

\subsection{Analyzing Cluster Color-Color Parameters and Selecting Turnoff Stars}

In Figures 5 and 6 we apply these color-color techniques to 12 young clusters ranging from the very 
young ($\sim$20 Myr) NGC 2547 to the intermediate-aged ($\sim$355 Myr) NGC 3532.  For all clusters
Gaia DR2 parallaxes and proper motions have been used to select likely cluster members.  The fit reddenings 
based on matching the F70 trend to the MSTO stars are labeled.  Our method fits the blue edge of the richest 
trend in the --0.1 $<$ (B--V)$_0$ $<$ 0.05 range and also considers the entirety of the reddening sensitive 
Hyades fiducial (orange line) fit of lower mass stars.  This avoids basing the fit on the more complex and 
sparser bluer stars.  In the clusters where differential reddening is 
important, focusing on the richest sequence of stars helps to define MSTO fits on the most common cluster 
reddening.  The final MSTO stars falling in our selected color-color range based on the F70 relation are 
plotted in red.  As a further reference in each cluster, we outline in cyan the previously identified Be 
stars and in blue the blue stragglers that are not Be stars.  Nearly all fall outside of the color-color 
selected range.

In Figures 5 and 6 the photometrically matched [Fe/H] (i.e., that applied to the Hyades fiducial fit) and 
the spectroscopic [Fe/H] are labeled for comparison.  They show reassuring consistency, or are within 
$\sim$0.1 dex, in all but one case.  The final cluster parameters and the sources of the spectroscopic 
[Fe/H] are all given in Table 3.  For clusters where spectroscopic [Fe/H] are taken from Cummings 
(2011) or Steinhauer \& Deliyannis (2004), their derivations of [Fe/H] are dependent on the adopted E(B--V) 
and B--V.  No adjustments were necessary for NGC 2168 from Steinhauer \& Deliyannis (2004), but for NGC 2547,
NGC 2422, and NGC 2516 from Cummings (2011) we have adjusted each cluster's spectroscopic [Fe/H] based on 
how our fit E(B--V) and observed B--V differ from that originally adopted.  For example, in the F and G 
dwarfs that were used in these sources a systematic decrease in (B--V)$_0$ of 0.01 results in an increase 
in spectroscopically derived [Fe/H] of $\sim$0.02 dex.

\begin{center}
\begin{deluxetable}{l c c c c}
\multicolumn{5}{c}%
{{\bfseries \tablename\ \thetable{} - Sources and Systematics}} \\
\hline
Cluster   & Phot    & $\Delta$(B--V) &$\Delta$(U--B) & $\Delta$V  \\
          & Sources & (Mag)          & (Mag)         & (Mag)      \\
\hline
NGC 2547  & 1,2     &  +0.030        & -             &  +0.045         \\
IC 2602   & 3,4     & \,\,--0.035    & -             & \,\,--0.065     \\
Alpha Per & 5,6     & \,\,--0.005    & -             & \,\,--0.050     \\
NGC 6405  & 7,8     & \,\,--0.020    & \,\,--0.030   &  +0.020         \\
NGC 2323  & 9,10    &  +0.015        & -             &  +0.010         \\
Pleiades  & 11      & -              & -             & -               \\
NGC 2422  & 12,13   &  +0.010        & +0.010        &  +0.080         \\
NGC 2516  & 14,15   &  +0.010        & +0.010        & \,\,\,\,\,0.000 \\
NGC 2168  & 16      & -              & -             & -               \\
NGC 2301  & 10,17   & \,\,\,\,\,0.000& -             & \,\,\,\,\,0.000 \\
NGC 2287  & 18,19   & \,\,--0.030    & -             & \,\,--0.050     \\
NGC 3532  & 20,21   &  +0.020        & -             & \,\,\,\,\,0.000 \\
\hline
\caption{If there are two photometric sources listed, the primary source is listed first and the secondary 
source for fainter stars and those not included in the primary source is listed second.  The listed $\Delta$
signs are based on the primary minus the secondary photometry.  (1) Claria (1982) (2) Naylor et~al.\ (2002) 
(3) Whiteoak (1961) (4) Prosser et~al.\ (1996) (5) Mitchell (1960) (6) Prosser (1992) (7) Eggen (1961) 
(8) Terzan et~al.\ (1987) (9) Claria et~al.\ 
(1998) (10) Kalirai et~al.\ (2003) (11) Johnson \& Mitchell (1958) (12) Hoag et~al.\ (1971) (13) Prisinzano 
et~al.\ (2003) (14) Dachs \& Kabus (1989) (15) Sung et~al.\ (2002) (16) Sung \& Bessell (1999) (17) Kim 
et~al.\ (2001) (18) Ianna et~al.\ (1987) (19) Sharma et~al.\ (2006) (20) Fernandez \& Salgado (1980) (21) 
Clem et~al.\ (2011).  }
\end{deluxetable}
\end{center}

\begin{center}
\begin{deluxetable}{l c c c c}
\multicolumn{5}{c}
{{\bfseries \tablename\ \thetable{} - The Effects of Magnitude Systematics}}\\
\hline
                     & $\Delta$E(B--V) &$\Delta$[Fe/H]$_{\rm p}$ & $\Delta$[Fe/H]$_{\rm s}$& $\Delta$(m--M)$_0$\\
                     &    (Mag)        & (dex)                   & (dex)                   & (Mag)             \\
\hline
$\Delta$U=+0.02 &  \,\,--0.005 & +0.10      & \,\,--0.01 & \,\,--0.03\\
$\Delta$B=+0.02 & +0.029       & \,\,--0.20 & +0.02      &  \,\,0.0\\
$\Delta$V=+0.02 & \,\,--0.025  & +0.10      & \,\,--0.01 &  +0.05\\
\hline
\caption{[Fe/H]$_{\rm p}$ represents photometrically derived [Fe/H] and [Fe/H]$_{\rm s}$ represents spectroscopically
derived [Fe/H].}
\end{deluxetable}
\end{center}

In Figure 7, we plot the (U--B)$_0$ residuals relative to the F70 relation for all Be (cyan), 
Ap (orange), and Am (yellow) stars identified from these 12 clusters (WEBDA and SIMBAD databases).  We also show 
in blue all MSTO stars that are blue stragglers but are not Be stars.  The color-color selection range is 
displayed, but we note that the curvature on one edge is due to how this range was designed to follow the curvature 
of the F70 relation.  Most Be stars in these clusters deviate from the F70 relation, and as expected they span from 
having either too much U flux or too little for a star at their B--V.  Note in Figure 7 that the two bluest 
Be stars (from NGC 2422 and NGC 2516) fall outside of the plotted range in Figure 6, but they are shown here to 
clearly deviate from the F70 relation.  Four of the Be stars, two from Alpha Per and two from NGC 2323, however, 
fall within this selected range.  This is consistent with Be stars going through phases of normal or near normal 
colors, and we will further consider these four in the CMDs in Section 4.   

More strikingly, the blue stragglers consistently show too little U flux in all cases with all falling well outside 
the selected range. We reiterate that these blue stragglers are all identified as cluster members based on 
Gaia DR2 data, so their colors are peculiar, but it is unclear why they have such red U--B colors relative to their B--V.   
We acknowledge that some of the blue stragglers with more moderate deviation may result from moderate differential reddening, but
overall this reassures us of their peculiarity and the strength of this color-color selection technique.  That
they all fall redward in U--B color also further distinguishes them from Be stars, which more commonly show U excess. 

Lastly, the peculiar abundance Am stars show minor scatter and Ap stars show larger scatter, but they predominantly fall 
within our selected range.  This is consistent with metallicity changes causing their colors to shift along the F70 relation 
rather than away from it at these colors.

\subsubsection{Systematic Photometric Errors}

For three clusters in Figures 5 and 6 we have combined two different data sets to increase the depth of their analysis.  
Based on the stars that were in both sets, we measured the systematics differences shown in Table 1.  Also shown in Table 1, 
for most other clusters we have matched their MSTO photometry to deeper BV photometry for the general CMD analysis (Section 4).  
Consistency between these combined photometric sets is important, and in all cases we adjust the deeper photometric data to 
match the zero point of the data set focusing on MSTO stars.  Besides increased depth, these sample comparisons 
also give us an estimate of the systematic zero-point errors for these photometric data sets.

\begin{center}
\begin{deluxetable*}{l c c c c c c c c c}
\multicolumn{10}{c}%
{{\bfseries \tablename\ \thetable{} - Open Cluster Parameters}} \\
\hline
Cluster   & E(B--V)$^a$     & [Fe/H]   & [Fe/H]  & Y$^2$ Age  & (m--M)$_0$    & PARSEC           & (m--M)$_0$    & MIST            &  SYCLIST\\
          &                 &          & Sources & (Myr)      &  Y$^2$        & (Myr)            &  PARSEC       & (Myr)           &   (Myr)  \\
\hline
NGC 2547  & 0.060$\pm$0.02  & +0.01          & 1 &  -         & 8.04$\pm$0.15 &   9$^{+15}_{-5}$ & 7.98$\pm$0.15 & 10$^{+15}_{-5}$ &  60$\pm$20 \\
IC 2602   & 0.050$\pm$0.02  &\,\,--0.02      & 2 &  -         & 5.90$\pm$0.15 &  10$^{+15}_{-5}$ & 5.84$\pm$0.15 & 16$^{+15}_{-5}$ &  60$\pm$20 \\
Alpha Per & 0.065$\pm$0.03  & +0.02          & 3 & 110$\pm$20 & 6.10$\pm$0.06 &  80$\pm$20       & 6.04$\pm$0.06 &  85$\pm$20      &  90$\pm$20\\
NGC 6405  & 0.120$\pm$0.03  & +0.04          & 4 & 120$\pm$20 & 8.48$\pm$0.06 & 100$\pm$20       & 8.42$\pm$0.06 & 105$\pm$20      & 110$\pm$20\\
NGC 2323  & 0.230$\pm$0.05  &\,\,\,\,\,0.00  & - & 140$\pm$35 & 9.92$\pm$0.10 & 115$\pm$35       & 9.86$\pm$0.10 & 125$\pm$35      & 120$\pm$35 \\
Pleiades  & 0.030$\pm$0.02  & +0.01          & 5 & 145$\pm$15 & 5.58$\pm$0.06 & 115$\pm$15       & 5.52$\pm$0.06 & 135$\pm$15      & 125$\pm$15 \\
NGC 2422  & 0.065$\pm$0.02  & +0.02          & 1 & 155$\pm$20 & 8.53$\pm$0.12 & 150$\pm$20       & 8.47$\pm$0.12 & 150$\pm$20      & 145$\pm$20 \\
NGC 2516  & 0.090$\pm$0.03  &\,\,\,\,\,0.00  & 1 & 195$\pm$25 & 8.11$\pm$0.12 & 165$\pm$25       & 8.04$\pm$0.12 & 195$\pm$25      & 185$\pm$25 \\
NGC 2168  & 0.240$\pm$0.05  &\,\,--0.143     & 6 & 195$\pm$30 & 9.58$\pm$0.10 & 175$\pm$30       & 9.52$\pm$0.10 & 180$\pm$30      & - \\
NGC 2287  & 0.030$\pm$0.02  &\,\,--0.11      & 2 & 200$\pm$25 & 9.18$\pm$0.08 & 200$\pm$25       & 9.11$\pm$0.08 & 200$\pm$25      & - \\
NGC 2301  & 0.040$\pm$0.03  &\,\,\,\,\,0.00  & - & 200$\pm$30 & 9.65$\pm$0.10 & 200$\pm$30       & 9.58$\pm$0.10 & 205$\pm$30      & 185$\pm$30\\
NGC 3532  & 0.030$\pm$0.02  &\,\,\,\,\,0.00  & 2 & 340$\pm$30 & 8.33$\pm$0.14 & 345$\pm$30       & 8.28$\pm$0.14 & 360$\pm$30      & 345$\pm$30\\
\hline
\caption{a) For reddenings we have adopted the color dependent reddening relation of Fernie (1963) and 
give the derived reddenings at a color of (B--V)$_0$=0.  We calculate true distance moduli based on 
extinctions of A$_V$=3.1$\times$E(B--V).  The spectroscopic sources are (1) Cummings (2011) (2) Netopil et~al.\ (2016) 
(3) Boesgaard et~al.\ (2003) (4) K{\i}l{\i}{\c c}o{\u g}lu et~al.\ (2016) (5) Schuler et~al.\ (2010) (6) Steinhauer \& Deliyannis (2004).  For spectroscopic 
[Fe/H] from Cummings (2011) and K{\i}l{\i}{\c c}o{\u g}lu et~al.\ (2016), we make appropriate adjustments based on updated stellar parameters.  For NGC 2323 
and NGC 2301, no spectroscopic [Fe/H] were available but their photometric [Fe/H] are consistent with 0.0.  
MIST and SYCLIST distance moduli are not given because they are indistinguishable from those of PARSEC and 
Y$^2$, respectively.}
\end{deluxetable*}
\end{center}

In Table 2 we consider what effects systematic shifts of +0.02 magnitudes in either U, B, or V will have 
on the color-color derived E(B--V).  For shifts in all three magnitudes, the resulting 
total effect on (B--V)$_0$ for cluster stars is weak because shifts in U weakly affect the fit E(B--V), and shifts 
in B or V result in comparable adjustments of B--V and the fit E(B--V).  Therefore, when stellar parameters are 
estimated using these (B--V)$_0$, the spectroscopic [Fe/H] are only weakly affected by these systematics.  For the 
photometric estimate of [Fe/H], however, these 0.02 magnitudes shifts will have much stronger effects ($\sim$0.10 
to 0.20 dex).  Lastly, after applying any of these systematic magnitude shifts to the photometry, in addition to the corresponding
adjustments to the E(B--V) and spectroscopic [Fe/H] to the isochrone, the extinction corrected true distance modulus 
((m--M)$_0$) is also found to be weakly sensitive to these systematic errors.  In all cases, these effects on E(B--V), 
[Fe/H], and (m--M)$_0$ scale approximately linearly with the color shifts.

The strong sensitivity of the photometric estimate of [Fe/H] to systematic photometry errors illustrates one of its 
limitations.  Reassuringly, all of the clusters analyzed here have both spectroscopic and photometric [Fe/H] in agreement, 
or within $\sim$0.1 dex.  Minor photometric systematics are a likely cause for any remaining [Fe/H] disagreements, but 
these systematics will have little to no effect on the derivation of the other parameters.  For the following CMD 
isochronal analyses, when available, each cluster's more robust spectroscopic [Fe/H] has been adopted.  

\section{Color-Magnitude Diagrams}

\begin{figure*}[!ht]
\begin{center}
\includegraphics[clip, scale=0.91]{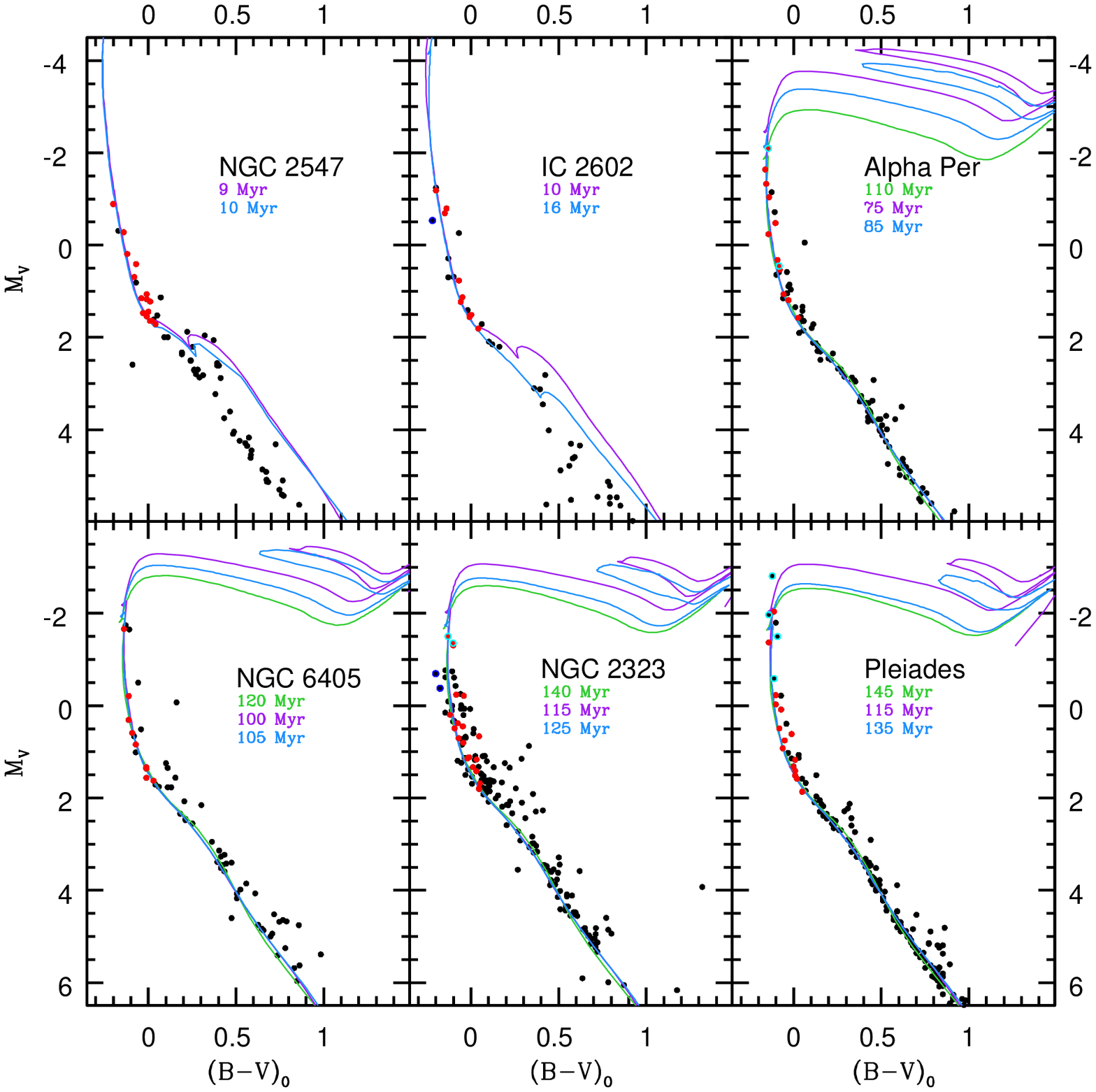}
\end{center}
\vspace{-0.4cm}
\caption{CMD analysis of the six youngest open clusters in our sample.  The cluster displayed in each
panel is labeled.  The photometric data have been corrected to place all clusters on their absolute
magnitude (M$_v$) and intrinsic (B--V)$_0$ scale.  Three isochrone fits are shown with the Y$^2$ 
isochrone (green), the PARSEC isochrone (purple), and the MIST non-rotating isochrone (blue).  
Again, Be stars are circled in cyan and blue stragglers that are not Be stars are circled in blue. 
For both the young NGC 2547 and IC 2602, the Y$^2$ isochrones do not consider high enough 
masses to analyze their MSTOs.  Also, note the disagreement below their MSTOs between the 
isochrones still on the pre-main sequences and their observed main sequences.  See Table 
1 and Table 3 for the photometric sources and measured cluster parameters, respectively.}
\end{figure*}

In young and relatively sparse cluster MSTOs with a lack of detailed rotation and inclination information, 
it is difficult to know which initial rotation rate to apply to the isochrones to best represent the 
MSTO stars.  Even in the cases where this information is partially available, the uncertainties in rotational 
models limit age accuracies.  The color-color techniques applied here, however, can also help mitigate these rotational 
MSTO challenges.  For example, Bastian et~al.\ (2017) found that in the rich Large Magellanic Cloud clusters 
NGC 1851 ($\sim$80 Myr) and NGC 1856 ($\sim$282 Myr), their MSTOs have a large fraction of Be stars with 
H$\alpha$ emission resulting from their circumstellar disks.  Estimates show that most B stars in a MSTO with 
rotations at or greater than 0.6 of $v_{crit}$ will be Be stars (Rivinius et~al.\ 2013 and references therein).  
Even though the color-color trends of rapidly rotating stars themselves are predicted to appear 
normal, the color-color analysis in Section 3.2 will select out active Be stars.  

\begin{figure*}[!ht]
\begin{center}
\includegraphics[clip, scale=0.91]{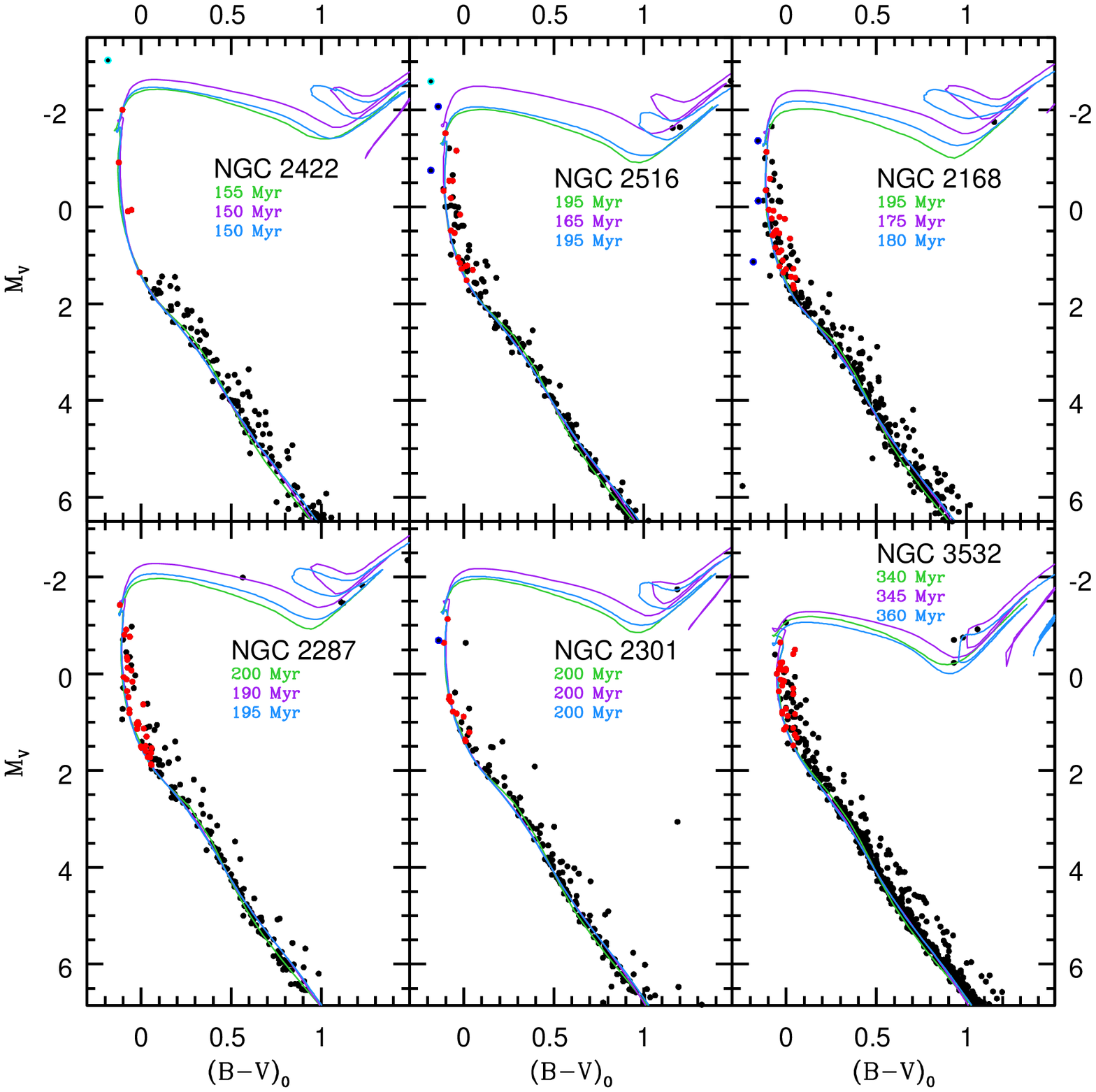}
\end{center}
\vspace{-0.4cm}
\caption{The same CMD analysis as shown in Figure 8 for the six older open clusters in our sample.}
\end{figure*}

In general, therefore, these color-color techniques will have selected out the fastest rotators through their 
disk's NUV flux.  These fastest rotators are also the stars most sensitive to angle of observation, and so their
removal limits the effects of that unknown parameter on the observed photometry.  The final color-color selected MSTO 
stars will be slow and intermediate velocity rotators, which have moderate sensitivity to rotation, as shown in 
Figure 1 and 2, and only a minor sensitivity to angle of observation.  That these should be expected to create 
a narrow MSTO is illustrated by the $\textit{v sin i}$ observed by Dupree et~al.\ (2017) in the MSTO stars in the 
200 Myr Large Magellanic Cloud cluster NGC 1866.  They found narrow line MSTO stars are preferentially bluer, broad 
line stars are preferentially redder, and Be stars are the most broadly distributed on the rich MSTO.  The rotational 
models, however, remain inconsistent in the details and this illustrates the remaining sensitivity to our understanding 
of rotation.  

\subsection{Color-Magnitude Diagram Isochrone Fitting}

The MIST isochrone rotational models (Figure 1) argue that at ages $\sim$100 Myr and 
older, the MSTO stars with intermediate rotational velocities are consistently redder than the slowest (non) 
rotators.  Therefore, MSTOs and their ages may best be found by fitting the blue edge of these color-color 
selected stars with non-rotating isochrones.  This suggests the Y$^2$ and PARSEC models, which do not consider 
rotation, also appropriately fit MSTO ages for those at $\sim$100 Myr and older.  This is similar to the 
technique adopted in Paper I, where we also fit the blue edge of the MSTO because this allows 
us to preferentially fit single stars and minimize the effects of binaries that are also common at these masses.  
The photometric effects of intermediate rotation rates, however, in Figure 1 suggest that in the youngest clusters
analyzed here (NGC 2547 and IC 2602), the non-rotating models may not provide accurate MSTO ages.  We will still, 
nonetheless, fit them with non-rotating isochrones and test for systematic issues in their ages relative to other 
cluster age techniques in Section 5.2.  

\begin{figure*}[!ht]
\begin{center}
\includegraphics[clip, scale=0.91]{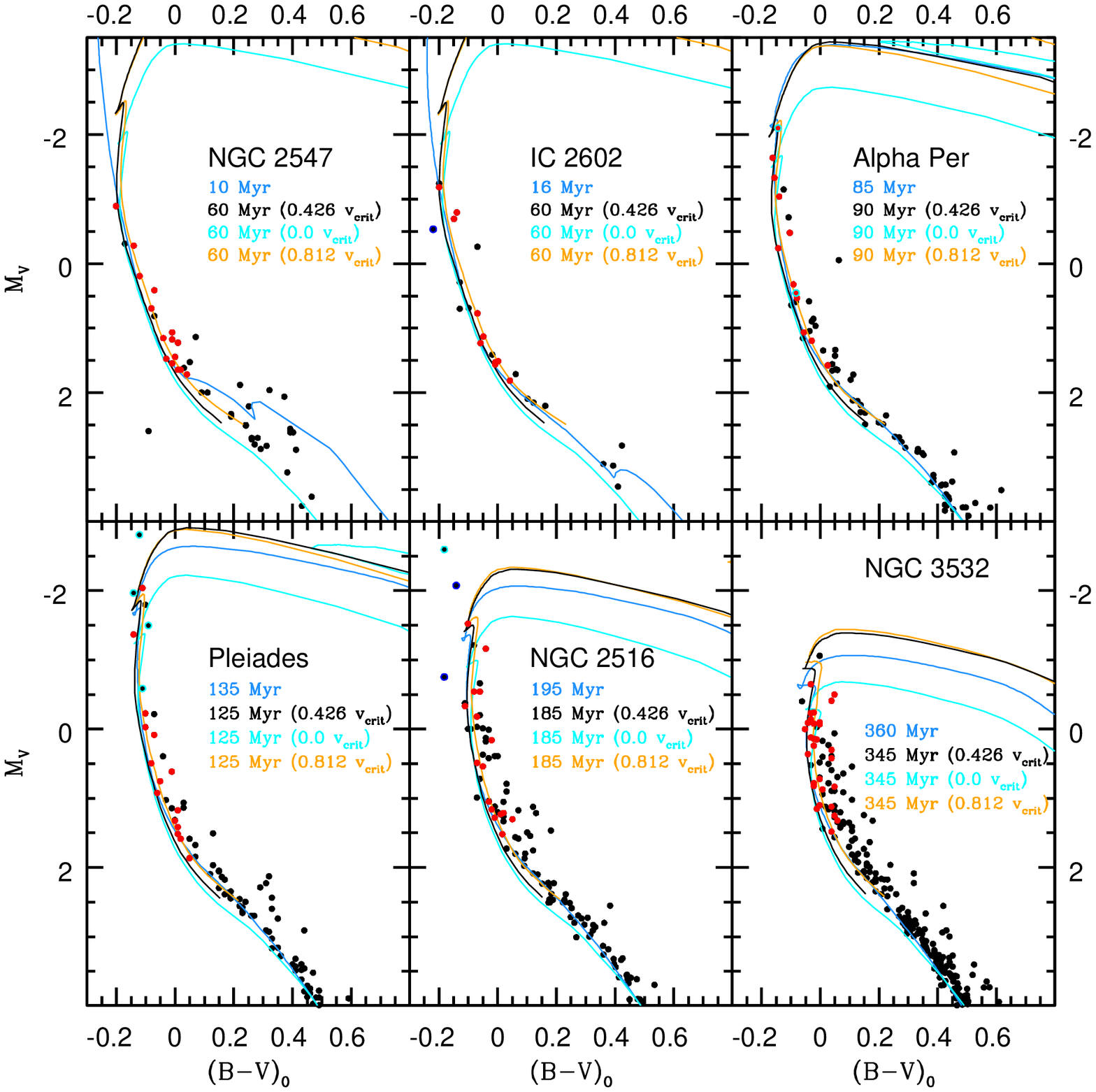}
\end{center}
\vspace{-0.4cm}
\caption{CMD analysis of six selected open clusters using the SYCLIST isochrones at $v_{crit}$ of 0.426
in black.  These are the same data from Figures 8 and 9, showing again the same MIST isochrone matches for
reference.  For each cluster we zoom in closer to the MSTOs.  This helps to illustrate the model's predicted 
effects of rotation on these MSTOs and the upper main sequences, where we show in comparison the SYCLIST
isochrones at the same ages but with no rotation (cyan) and at $v_{crit}$ of 0.812 (orange).  See Table 3 
for the measured cluster parameters.}
\end{figure*}

In Figures 8 and 9 we show the CMDs for these 12 clusters.  For six of these clusters, these are 
updated fits of those analyzed in Paper I.  The parameters are fit with the Y$^2$ isochrones (green), the 
updated PARSEC isochrones (purple), and the non-rotating MIST isochrones (blue).  The data points colored 
in red illustrate the same MSTO stars that have been selected with the color-color technique shown in Figures 
5 and 6.  Again, only likely cluster members based on Gaia DR2 parallaxes and proper motions are shown.
For several clusters that are in rich Galactic fields, a few deviant stars that are likely non-members remain, 
but this shows that non-member contamination is very small and is not a concern in our analysis. 
For all isochrones the distance moduli are independently fit considering both the main sequence and the selected MSTO 
stars, but the distance moduli fit by the PARSEC and MIST models were indistinguishable. 

To provide uniform MSTO age analysis, for each isochronal model we have fit the blue edge of the selected 
sample of MSTO stars.  We used our own program that takes a grid of input isochrones at different ages and matches the best 
fitting isochrone age through orthogonal regression.  This is a least squares technique that considers separation along 
both the absolute magnitude and color axes.  However, when looking at an isochrone across a range of ages, the lower MSTO 
changes slowly with age while the upper MSTO changes very rapidly with age.  Therefore, our program also measures each star's 
sensitivity to age variations, relative to this input grid of isochrones, and weights its least squares input.
For the youngest clusters ($<$20 Myr) we consider age increments of 1 Myr, and for the older clusters ($>$50 Myr)
we consider age increments of 5 Myr.

For all clusters the color-color selected stars create better defined MSTO trends.  These selections have 
also removed all clear blue stragglers (circled in blue) from the clusters.  Additionally, in all but the 
sparsest clusters the distribution of selected MSTO stars show a consistent and well defined blue edge to the MSTO that 
matches well with the full extent of all three non-rotating isochronal models.  These selected MSTOs are 
narrower but do remain too broad to simply be explained by binaries.  We also note that the Be stars (circled 
in cyan) typically deviate significantly from the MSTO.  However, as discussed in Section 3.3, four of these 
Be stars have color-colors consistent with our selection (see Alpha Per and NGC 2323).  In these two cluster CMDs 
all of these Be stars are photometrically consistent with their cluster's other selected MSTO stars and their 
photometry has marginal to no affect on our age fits.  Overall, this may indicate that these Be stars have, or 
were photometrically observed during a phase of, relatively normal colors.

While both the MIST and PARSEC isochrones can also well match the blue edge of the MSTOs of the very young NGC 
2547 and IC 2602, the limitations discussed above of using non-rotating isochrones to analyze ages of such young 
clusters is apparent.  Below their MSTOs there is clear disagreement between the observed main sequence 
stars and each isochrone's predicted pre-main sequence extending nearly to the MSTO.  This will be discussed further
in the Section 4.1.1. 

Lastly, while all three isochrones consistently fit the shape of the MSTO stars and have similar main
sequence trends, we note that their post main sequences have larger differences.  Based on this and the limited
number of giants in the older clusters here, which makes it difficult to define their trends, we have not adjusted 
our isochrone fits based on these giants.  Nevertheless, we note the encouraging consistency between these 
independently fit isochrones and the giants in NGC 2516, NGC 2168, NGC 2301, NGC 2287, and NGC 3532.  

In Table 3 we list for each cluster the derived reddening, the adopted metallicity, and for each 
isochrone the derived age and distance modulus.  The photometric sources are given in Table 1.  In Table 3
we also give the source of the adopted [Fe/H].  

\subsubsection{SYCLIST Isochrone Fits}

In Figure 10 for select clusters with solar, or nearly solar, metallicity we also fit these CMDs using SYCLIST 
isochrones.  For matching to the intermediate main sequence (0.5 $<$ (B--V)$_0$ 
$<$ 1.0; region not shown in Figure 10) there is no meaningful difference between the non-rotating 
isochrones and those at 0.4 of $v_{crit}$.  This is an important consideration because the MIST isochrones do not 
model rotation at these lower masses, and this suggests that the intermediate main sequence photometry that is
valuable to the distance modulus measurements is not meaningfully affected by rotation.  We also note that the 
distance moduli required to match the SYCLIST isochrones with the observations for all clusters are found 
to be indistinguishable from those used to match the Y$^2$ isochrones.

Matching the SYCLIST models at the intermediate main sequence finds that, unlike all other models, the 
non-rotating SYCLIST isochrone (cyan) provides a typically poor match to the upper main sequence (0.06 $<$ (B--V)$_0$ 
$<$ 0.3) and to the selected MSTO, and in a manner where adjusting the age would not help.  To match the observed 
photometry in the upper main sequence and the faint MSTO stars requires very rapid initial rotations of 
0.95 of $\Omega_{crit}$ (0.812 $v_{crit}$; orange).  This velocity is abnormally large.  We note that Royer et~al.\ 
(2007) have found that the A dwarfs in this region are typically rotating at a higher fraction of their 
$v_{crit}$ than B dwarfs, but it is unlikely that most of these selected faint MSTO stars and the A dwarfs 
are rotating at near critical velocity.

Based on typical B dwarf rotations from Huang et~al.\ (2010) and the rotation sensitivity of the SYCLIST models 
shown in Figure 2, with the orthogonal minimization techniques discussed above we have fit the blue edge of these 
selected MSTO stars using SYCLIST isochrones with initial rotations of 0.6 of $\Omega_{crit}$ (0.426 of $v_{crit}$).  
While these isochrones remain too faint at the upper main sequence and faintest MSTO stars, this rotation
rate otherwise provides convincing fits of the blue edge of the MSTO.  

For the sub-sample of clusters analyzed using SYCLIST models, we also give their SYCLIST-based ages in Table 3.
These moderately rotating SYCLIST models fit comparable ages to the non-rotating MIST, Y$^2$, and PARSEC isochrones 
for the clusters at $\sim$100 Myr and older.  However, the SYCLIST models fit ages that 
are increasingly older relative to these non-rotating isochrones in the youngest clusters ($<$100 Myr) analyzed 
here.  These differences are clear in the young NGC 2547 and IC 2602, where as seen in Figure 8
these non-rotating isochrones fit MSTO ages that would place nearly all of the lower-mass stars below the MSTO
on the pre-main sequence.  In contrast to this, these significantly older cluster ages found with rotating 
SYCLIST models do predict main sequences consistent with observations.

\section{Remaining Challenges \& Comparisons}

These color-color techniques, use of multiple isochrones, Gaia DR2 memberships, and consistent analysis 
across a broad range of cluster parameters have helped improve cluster MSTO age analysis.  For the clusters at 
$\sim$100 Myr and older there is strong consistency of the age fits across all four isochrones, but there remain 
several important challenges.  These include the broad differences between the SYCLIST models, 
including their non-rotating models, and all other isochrones.  For example, at $>$100 Myr the MIST 
models (Figure 1) predict that the blue edge of the MSTO is defined by slowly (non) rotating stars but the 
SYCLIST models (Figure 2) predict that the blue edge of the MSTO is defined by moderately rotating ($\sim$0.426 
$v_{crit}$) stars.  Even though adopting these MSTO characteristics for the corresponding model results 
in consistent ages, both assumptions cannot be correct.  

Comparisons to cluster parameters derived with independent methods, in particular those insensitive to
many of the challenges discussed above, are valuable.  These provide references for these general analysis
techniques and for the differences found between different isochrones.  This is one of the main advantages
of testing these methods relative to nearby clusters that can be analyzed using a number of methods.

\subsection{Comparison to Gaia-Based Distances}

In addition to our membership analysis, Gaia DR2 provides a valuable check of distance moduli for nearby 
open clusters (Gaia Collaboration et~al.\ 2018b).\footnote{We correct these cluster parallaxes for the Gaia 
DR2 systematic parallax error of 0.029 mas (Lindegren et~al.\ 2018).}  These include ten of the clusters analyzed here, in addition to 
the Hyades and Praesepe analyzed similarly in Cummings et~al.\ (2017), and NGC 1039 (M39) and NGC 6475 (M7) analyzed 
for this comparison using the same techniques applied in this paper.  The photometric determination of distance 
moduli are dependent on multiple factors including the photometric UBV zero points, reddening determination, adopted 
composition, extinction ratio, photometric fitting errors, and adopted isochrones.   Figure 11 and Table 4 show the 
direct comparison of Gaia-based distance moduli to the distance moduli derived from our Y$^2$ and SYCLIST fits.  We 
calculated the photometric-based true distance moduli by adopting an extinction correction of A$_V$=3.1$\times$E(B--V).  
For all 14 clusters there is strong consistency (mean differential of --0.01 with a $\sigma$ of 0.10 magnitudes), 
reassuring us that the determination of reddening, adopted composition, and the general techniques applied here are 
reliable.

\begin{figure}[!ht]
\begin{center}
\includegraphics[clip, scale=0.44]{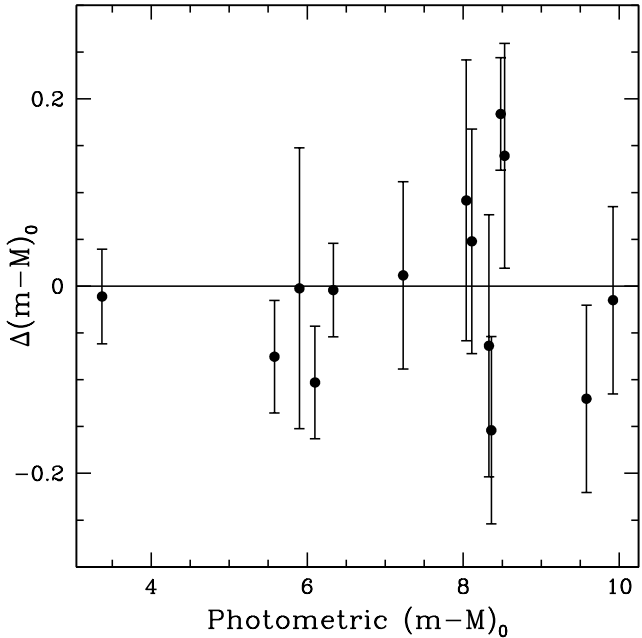}
\end{center}
\vspace{-0.4cm}
\caption{The difference in true distance modulus derived from the direct photometric fits using the
Y$^2$ and SYCLIST isochrones minus those derived from Gaia DR2 mean cluster parallaxes.  The published
gaia distance errors are added in quadrature to the photometric distance modulus errors,
but the parallax errors are of order the size of the data points.}
\end{figure}

Lastly, the distance moduli fit with PARSEC and MIST isochrones are 0.06 to 0.08 magnitudes lower than those 
determined with the Y$^2$ and SYCLIST isochrones.  Comparing the smaller PARSEC/MIST distance moduli for these 
14 clusters results in a larger (--0.08$\pm$0.10 magnitudes) difference from the Gaia distances.  

\subsection{Comparison to Lithium Depletion Boundary Age Techniques}

These MSTO derived ages can also be compared to cluster ages determined from Li abundances of low-mass cluster 
members (LDB ages; e.g., Jeffries \& Naylor 2001).  Dahm (2015), for example, has analyzed the lowest-mass stars 
in the Pleiades to identify the age sensitive luminosity where Li abundances rapidly change from fully Li depleted to Li rich.  
LDB ages provide a valuable age reference and are believed to be relatively insensitive to model assumptions like 
opacities, equations of state, metallicity, and rotation (Soderblom et~al.\ 2014), but they are not completely 
independent of models (Tognelli et~al.\ 2015).  Based on differing pre-main sequence evolutionary models, for 
example, Dahm (2015) derives 3 Pleiades ages of 116 Myr (using Baraffe et~al.\ 1998 models), 108 Myr (using Baraffe 
et~al.\ 2015 models), and 103 Myr (using Chabrier et~al.\ 2000 models).  These ages are comparable to our MSTO 
age fits, where the PARSEC, MIST, and SYCLIST ages span a similar range of ages, but the LDB ages are 
approximately 15 Myr younger.  

One challenge with LDB ages is that while the typical rotations in low-mass stars are not believed to 
play a direct role, in these stars with surface convection zones the rotation correlates with surface 
activity and results in radius inflation through induced magnetic fields (Somers \& Stassun 2017).  Therefore, 
low-mass stars that are initially rapidly rotating undergo less Li depletion because this inflation decreases 
their interior temperatures (Somers \& Pinsonneault 2014).  Additionally, this activity also affects a star's 
photometric parameters at a given mass through inflation causing lower T$_{\rm eff}$ and a rich coverage of 
starspots on their surface (Somers \& Pinsonneault 2015).  These factors are complex, however, and we need to 
better understand typical starspot coverage percentages.  Accounting only for the photometric effects will
decrease the standard LDB age by $\sim$10\% (Juarez et~al.\ 2014).  By factoring in both the effects on 
photometry and Li depletion rate, Jackson \& Jeffries (2014) and Somers \& Pinsonneault (2015) predict that 
at the age of the Pleiades this will increase the standard LDB ages by as much as $\sim$20\% and 10\%, 
respectively.  This can explain the systematic 15 Myr difference between the standard LDB ages and MSTO 
ages for the Pleiades.

In Table 4 the MSTO fit ages based on the PARSEC and SYCLIST isochrones are compared to the published LDB ages 
that have been corrected based on Jackson \& Jeffries (2014) at 30\% starspot coverage.  All four isochrones derive quite similar ages 
for older clusters, but the differences become more significant in the youngest clusters analyzed here: NGC 2547 
and IC 2602.  The PARSEC and MIST non-rotating isochrones derive 9 and 10 Myr, respectively, for NGC 2547 and 10 
and 16 Myr, respectively, for IC 2602.  Our SYCLIST fits derive 60 Myr for NGC 2547 and 60 Myr for IC 2602.  LDB 
ages for NGC 2547 and IC 2602 have been measured at 35 Myr (Jeffries \& Oliveira 2005) and at 46 Myr (Dobbie, 
Lodieu \& Sharp 2010), respectively.  These LDB ages adopted standard models without rotation and activity.  This 
gives LDB ages adjusted for activity of 45 Myr for NGC 2547 and of 59 Myr for IC 2602.  These are significantly older 
than the MSTO ages fit using either the PARSEC or MIST isochrones, but they are consistent with the SYCLIST-based ages.

The cluster Alpha Per also has an LDB age determination of 90$\pm$10 Myr (Stauffer et~al.\ 1999).  Adjusting for activity gives 
107 Myr.  Comparison to our Alpha Per MSTO ages again shows systematically younger ages based on the PARSEC and MIST 
isochrones and marginally better consistency with the SYCLIST models.  Lastly, Mart{\'{\i}}n et~al.\ (2018) have 
measured an LDB age of 650$\pm$70 Myr for the Hyades from its low mass L dwarfs.  Neither Jackson \& Jeffries (2014) nor Somers \& 
Pinsonneault (2015) model the effects of rotation and magnetic fields at these older ages, but L dwarfs are less 
sensitive to these effects and any corrections are likely minor.  This LDB age is consistent with the MSTO age fit in 
Cummings et~al.\ (2017) using the Y$^2$ isochrones of 635 Myr.  Additionally, from Cummings et~al.\ (in prep.)
we have fit the Hyades with the PARSEC and MIST isochrones at 700 and 695 Myr, respectively.  There are no available metal-rich 
SYCLIST isochrones, however, so we have not also fit the Hyades with these rotating models.

\begin{figure}[!ht]
\begin{center}
\includegraphics[clip, scale=0.44]{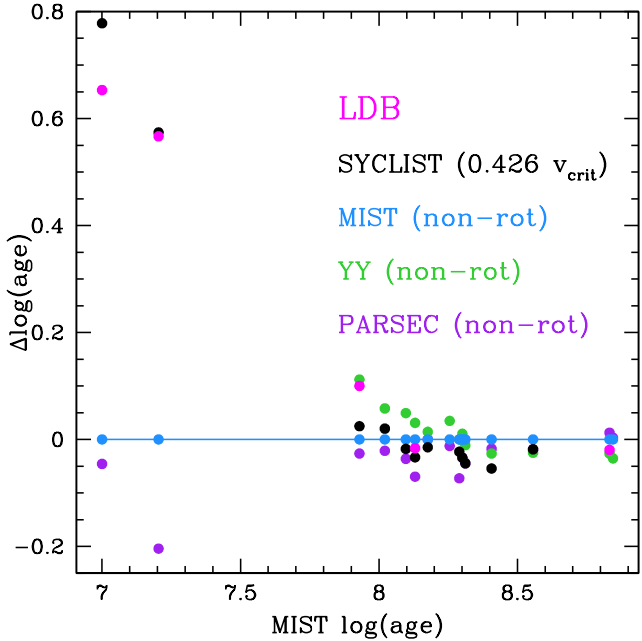}
\end{center}
\vspace{-0.4cm}
\caption{The comparisons of age methods discussed here, relative to the age determined from the non-rotating 
MIST isochrones.  This looks at clusters ranging from the youngest at NGC 2547 and IC 2602 to the intermediate-aged Hyades 
and Praesepe.  The isochronal ages are represented by the same colors adopted in Figures 8, 9, and 10, and the 
LDB ages are represented in magenta.  The consistency for all methods is illustrated for clusters at $\sim$100 
Myr and older.  In younger clusters the ages determined with the rotating SYCLIST isochrones are still consistent
with LDB ages, but they quickly diverge from the non-rotating PARSEC and MIST isochronal age fits.}
\end{figure}

Overall, the clusters analyzed with these 
isochrones at $>$100 Myr provide ages reassuringly consistent with activity corrected 
LDB ages.  In the youngest clusters, the rotating SYCLIST models remain consistent with the 
activity corrected LDB ages, but the non-rotating PARSEC and MIST isochrones increasingly underestimate 
these youngest cluster ages.  In Figure 12 we illustrate all of these age comparisons.

\begin{center}
\begin{deluxetable*}{l c c c c c c}
\multicolumn{7}{c}%
{{\bfseries \tablename\ \thetable{} - Comparison to Gaia Distances and LDB Ages}} \\
\hline
Cluster   & MIST Age & (m--M)$_0$ & SYCLIST Age & (m--M)$_0$  & LDB Age & (m--M)$_0$\\
          & (Myr)      &  (MIST)  & (Myr)       & (SYCLIST)   & (Myr)   &  Gaia    \\
\hline
NGC 2547  &  10$^{+15}_{-5}$& 7.98$\pm$0.15 &  60$\pm$20 & 8.04$\pm$0.15 & 45  & 7.973$\pm$0.001\\
IC 2602   &  16$^{+15}_{-5}$& 5.84$\pm$0.15 &  60$\pm$20 & 5.90$\pm$0.15 & 59  & 5.912$\pm$0.002\\
Alpha Per &  85$\pm$20      & 6.04$\pm$0.06 &  90$\pm$20 & 6.10$\pm$0.06 & 107 & 6.214$\pm$0.002\\
NGC 6405  & 105$\pm$20      & 8.42$\pm$0.06 & 110$\pm$20 & 8.48$\pm$0.06 &  -  & 8.325$\pm$0.003\\
NGC 2323  & 125$\pm$35      & 9.86$\pm$0.10 & 120$\pm$35 & 9.92$\pm$0.10 &  -  & 9.997$\pm$0.004\\
Pleiades  & 135$\pm$15      & 5.52$\pm$0.06 & 125$\pm$15 & 5.58$\pm$0.06 & 130 & 5.664$\pm$0.002\\
NGC 2422  & 150$\pm$20      & 8.47$\pm$0.12 & 145$\pm$20 & 8.53$\pm$0.12 &  -  & 8.421$\pm$0.002\\
NGC 2516  & 195$\pm$25      & 8.04$\pm$0.12 & 185$\pm$25 & 8.11$\pm$0.12 &  -  & 8.088$\pm$0.001\\
NGC 2168  & 180$\pm$30      & 9.52$\pm$0.10 &  -         & 9.58$\pm$0.10 &  -  & 9.756$\pm$0.003\\
NGC 1039  & 200$\pm$20      & 8.30$\pm$0.10 & 185$\pm$20 & 8.36$\pm$0.10 & -   & 8.546$\pm$0.003\\
NGC 6475  & 255$\pm$20      & 7.17$\pm$0.10 & 225$\pm$20 & 7.23$\pm$0.10 & -   & 7.236$\pm$0.001\\
NGC 3532  & 360$\pm$20      & 8.28$\pm$0.14 & 345$\pm$20 & 8.33$\pm$0.14 & -   & 8.424$\pm$0.001\\
Hyades    & 680$\pm$25      & 3.33$\pm$0.05 &  -         & 3.37$\pm$0.05 & 650 & 3.384$\pm$0.007\\
Praesepe  & 700$\pm$25      & 6.29$\pm$0.05 &  -         & 6.33$\pm$0.05 & -   & 6.350$\pm$0.001\\
\hline
\end{deluxetable*}
\end{center}
\vspace{0.5cm}

\section{Summary \& Conclusions}

Deriving MSTO ages in young clusters can be affected by binarity, rapid rotation, circumstellar disks, 
chemical peculiarities, blue stragglers, and differential reddening.  Detailed spectroscopic analysis, 
or narrowband photometry, of the MSTO stars can help identify some types of peculiar stars.  These are time 
consuming, however, and the effects of disks in Be stars are variable.  Broadband UBV color-color analysis 
of MSTO stars is more direct and their color-colors are measurably affected by 1) stars with companions 
that significantly affect their colors, 2) stars affected by differential reddening, 3) rapidly rotating Be 
stars, and 4) blue stragglers.  With this analysis the nature of the peculiarity that causes such a star to 
deviate cannot be determined, but this is not the goal.  This method instead works as a simple and simultaneous 
way to identify well behaved MSTO stars that provide higher precision MSTO ages.

These selected MSTO stars create narrower MSTOs with blue edges that are consistently well fit by non-rotating 
models from Y$^2$, PARSEC, and MIST.  The predicted effects of rotation on the MSTOs are dependent on the 
rotational models, in addition to the effects of observation angle, but the fastest rotating MSTO stars have 
been selected out through their disks (Be status) and resulting peculiar U flux.  The remaining slow and 
intermediate rotators are less sensitive to the effects of rotation and observation angle.  Most of the widths 
of these narrower selected MSTOs can likely be explained by the remaining rotation rates, differences in 
observation angle, and binaries and chemical peculiar stars that these color-color techniques were insensitive 
to.  Small age spreads within clusters may still be necessary to explain these remaining widths (e.g., Niederhofer 
et~al.\ 2015), but a better understanding of rotation's effects on MSTO photometry is necessary to quantify this.

For the case of the SYCLIST isochrones, in contrast to the other models, their non-rotating isochrones 
provide poor fits to the cluster MSTOs.  In general, moderate to fast rotating SYCLIST isochrones are necessary to match the 
MSTO, and those at 0.426 of $v_{crit}$ are used to fit cluster ages.  For clusters older than 100 Myr, 
these SYCLIST age fits are reassuringly consistent with the non-rotating Y$^2$, PARSEC, and MIST model 
fits, even though the methods of addressing rotation are fundamentally different.  This
is consistent with the recent findings of Gossage et~al.\ (2018), where they created synthetic MSTOs 
based on variations in rotation rate and angle of observation and found the resulting ages for the 
Hyades, Praesepe, and the Pleiades are insensitive to the effects of rotation.  For younger 
clusters, however, where the importance of rotation increases, the SYCLIST fit ages become significantly 
older than those found with non-rotating isochrones.

The fit distance moduli of these clusters are important checks of our derived parameters.  Comparisons of 
the true distance moduli based on Y$^2$ and SYCLIST isochrone fits to the distance moduli derived from Gaia 
DR2 parallaxes show good agreement (--0.01$\pm$0.10 magnitudes).  This consistency is strong and shows that
with a larger sample of clusters these comparisons can discriminate between the distance moduli 
derived with the Y$^2$ and SYCLIST isochrones versus those from the PARSEC or MIST isochrones that are 
systematically closer (--0.08$\pm$0.10 magnitudes relative to Gaia).  This would provide a valuable 
constraint on how main sequence luminosity and color depend on mass and metallicity.

LDB ages provide a valuable independent check of MSTO cluster ages.  Taken from various sources, we have 
made corrections to these based on magnetic surface activity estimates and the resulting radius inflation.  
At the age of the Pleiades ($\sim$125 Myr), this gives consistency between our MSTO ages from differing 
isochrones and LDB ages from differing models.  The LDB ages of the much older Hyades ($\sim$675 Myr) also
remain consistent with our isochronal ages.

NGC 2547 and IC 2602, the youngest clusters we have analyzed, also have LDB ages and these provide an important
check between the non-rotating PARSEC and MIST isochrones fits (ranging from 9 to 16 Myr) and the LDB ages based 
on surface activity corrections (45 and 59 Myr, respectively).  Fitting these clusters with rotating MIST models at 0.5 or 0.6 of 
$v_{crit}$ would moderately increase the cluster age fits (by $\sim$5 to 10 Myr) but not enough to bring them to 
agreement with the LDB ages.  In contrast to this, fitting SYCLIST isochrones with initial rotations of 0.426 of 
$v_{crit}$ to the MSTO provides isochronal ages (60 Myr) reassuringly consistent with their LDB ages.  The moderately
older Alpha Per ($\sim$90 Myr) has all isochronal ages still generally consistent, but shows evidence of 
the non-rotating PARSEC and MIST ages beginning to diverge to younger fit ages than the SYCLIST and LDB ages.

Further work is necessary on evolutionary models and rotation, but here we have studied from several 
angles some of the challenges introduced by rotation and peculiar stars.  These simple Johnson UBV 
color-color techniques, which should be just as applicable in other broadband systems with a NUV filter, 
help address the most significant rotational challenges and peculiarities.  This improves the 
precision of cluster MSTO fitting.  More work on observational constraints of rotation will be required to 
improve these models.  For example, surface abundances in fast rotating stars and how they can trace 
rotational mixing (e.g., Boron, Venn et~al.\ 2002, Profitt et~al.\ 2016).  Astroseismology of rapidly 
rotating stars can also help characterize the effects of rapid rotation on interior structure (e.g., 
Neiner et~al.\ 2012).  Additional techniques to observationally constrain rotational mixing are valuable, 
however, and our group is currently developing a novel technique.

\vspace{0.5cm}
Acknowledgments: 
This project was supported by the National Science Foundation (NSF) through grant AST-1614933.
This research has made use of the WEBDA database, operated at the Department of Theoretical Physics 
and Astrophysics of the Masaryk University.  This research has made use of the SIMBAD database,
operated at CDS, Strasbourg, France.  JDC would also like to thank Constantine Deliyannis and Jieun
Choi for discussions on several of the ideas presented in this paper.

This work has made use of data from the European Space Agency (ESA) mission
{\it Gaia} (\url{https://www.cosmos.esa.int/gaia}), processed by the {\it Gaia}
Data Processing and Analysis Consortium (DPAC,
https://www.cosmos.esa.int/web/gaia/dpac /consortium). Funding for the DPAC
has been provided by national institutions, in particular the institutions
participating in the {\it Gaia} Multilateral Agreement.

\end{document}